\documentclass[a4paper,fleqn,usenatbib]{mnras}

\usepackage{mathptmx}

\usepackage[T1]{fontenc}
\usepackage{ae,aecompl}

\usepackage{graphicx}	
\usepackage{amsmath}	
\usepackage{amssymb}	

\title[Instabilities of non-stationary radiative shocks]{On physical and numerical instabilities arising in simulations of non-stationary radiatively cooling shocks}

\author[D. A. Badjin et al.]{
D. A. Badjin,$^{1,2}$\thanks{E-mail: badjinda@gmail.com}
S. I. Glazyrin,$^{1,2,3}$
K. V. Manukovskiy$^{2}$
and S. I. Blinnikov$^{2,1,4}$
\\
$^{1}$Centre for Fundamental and Applied Research, Dukhov Research Institute of Automatics (VNIIA), \\
  22, Sushchevskaya St,  Moscow 127055, Russia\\
$^{2}$Institute for Theoretical and Experimental Physics (ITEP), NRC `Kurchatov Institute', \\
   25, Bolshaya Cheremushkinskaya St, Moscow 117218, Russia\\
$^{3}$National Research  Nuclear University MEPhI, 31, Kashirskoye hwy, Moscow, 115409, Russia \\
$^{4}$Kavli Institute of Physics and Mathematics of Universe (IPMU, WPI), Tokyo University,\\
5-1-5 Kashiwanoha,  Kashiwa, 277-8583, Japan}

\date{Accepted XXX. Received YYY; in original form ZZZ}

\pubyear{2015}

\begin{document}
\label{firstpage}
\pagerange{\pageref{firstpage}--\pageref{lastpage}}
\maketitle

\begin{abstract}
We describe our modelling of the radiatively cooling shocks and their thin shells with various numerical tools in different physical and calculational setups. 
We inspect structure of the dense shell, its formation and evolution, pointing out physical and numerical factors that sustain its shape and also may lead to instabilities.
We have found that under certain physical conditions, the circular shaped shells show a strong bending instability and successive fragmentation on Cartesian grids soon after their formation, 
while remain almost unperturbed when simulated on polar meshes. We explain this by physical Rayleigh--Taylor like instabilities triggered by corrugation of the dense shell surfaces by numerical noise.   
Conditions for these instabilities follow from both the shell structure itself and from episodes of transient acceleration during re-establishing of dynamical pressure balance after sudden radiative cooling onset.
They are also easily excited
by physical perturbations of the ambient medium.
The widely mentioned Non-linear Thin Shell Instability, in contrast, in tests with physical perturbations is shown to have only limited chances to develop in real radiative shocks, as it seems to require a special spatial 
arrangement of fluctuations to be excited efficiently. 
The described phenomena also set new requirements on further simulations of the radiatively cooling shocks in order to be physically correct and free of numerical artefacts.
\end{abstract}

\begin{keywords}
supernovae: general -- ISM: supernova remnants -- hydrodynamics -- instabilities -- methods: numerical 
\end{keywords}

\section{Introduction}

Shocks in regime of radiative cooling or, more generally, volumetric energy losses, can be found in various astrophysical 
objects like young jet emitting stars \citep{,YSOHartigan2003,PlutoRadCool} or relativistic outflows of gamma-ray bursts  \citep[see][for the most recent review]{KumarZhang},
but probably the most widely mentioned are radiatively cooling shocks (hereafter RCSs) in supernova remnants (SNR) which lose
their energy due to photon emission and cosmic ray acceleration 
\citep{Loz1992sswi.book,MalkovDrury2001,Cox2005,
Vink2012,Zirakashvili2008,Slane2015}.  The volumetric losses effectively increase compressibility of matter, so that the most of the shocked fluid  
may collapse into a thin dense shell \citep{Cox1972, 
Falle1975,Falle1981,Cioffi1988,BKSilich1995,Chevalier1999,Petruk2006,Hnatyk2007}.

Dense shock-bounded shells are known to be subjected to various instabilities from quite general Rayleigh--Taylor and 
Richtmyer--Meshkov (RT and RM, respectively) to specific ones like linear and non-linear instabilities originally 
described by \citet{Vishniac1983, VishniacRyu89, Vishniac1994}. Numerous plasma instabilities of electrodynamical origin also should be
mentioned, however in this paper we restrict ourselves with pure hydrodynamics and non-stationary blast waves experiencing transition from adiabatic expansion into the radiative cooling regime. 

Development of the instabilities may result in a progressive deformation of the shell followed by
its subsequent fragmentation, enhancement of material mixing, amplification and dissipation of chaotic magnetic fields. 
E.g. in SNRs these phenomena might affect its net efficiency in cosmic ray acceleration and in deposition of momentum and energy 
into the interstellar medium.

Our group has come to studies of the RCSs while developing numerical tools of multidimensional
radiative hydrodynamics which are to be applied to models of strong and brightly shining shocks of superluminous supernovae
(SLSNe) under the Dense Shell Method (DSM) paradigm \citep{BaklanovEtAl_JETP_Lett_2013}. While carrying out test runs in the
optically thin radiative cooling regime, we noticed a strong instability of the RCS thin shell which displayed itself in vigorous 
short-scale bending with typical wavelength $\lambda$ of several shell thicknesses $h$ and amplitudes $\Delta R$
of several $\lambda$ ($h,\,\lambda\ll R_{\text{sh}}$, $\Delta R\sim 0.1R_{\text{sh}}$). It appeared even in cases of spherical-symmetric 
explosions in a uniform medium, i.e. without deliberately superimposed deviations, suggesting a numerical origin of
the instability or at least of its seminal perturbations, which then might be amplified by physical mechanisms inherent in equations being solved.

Some authors who modelled shocks with volumetric losses reported similar bending effects to appear in their 
simulations \citep{SchureKosenko09,KimOstriker14}, while others did not \citep[][ strictly speaking, they did as well, but their observed perturbations were much weaker]{vanMarle10}. 
As the two main distinctive features of the latter paper 
were in use of polar grid and an exact implicit cooling term accounting suggested by \citet{Townsend_ApJS_2009}, we decided to test different hypotheses of 
the instability generation: by mesh perturbations,  by inexact cooling treatment or by some possible artefacts of a certain code. 

To better understand conditions for possible instabilities in the simulated blast waves we have also undertaken a detailed analysis of high-resolution one-dimensional simulations.

In this paper we are going to show that bending and fragmentation, obtained in simulations, 
result from an interplay of numerical noise superimposed by rectangular grids
on to curved flows and physical instabilities of the two-shock bounded cold dense shell. 
There are fulfilled the conditions required for a non-linear thin shell instability \citep[NTSI][]{Vishniac1994} 
Rayleigh--Taylor and Richtmyer--Meshkov instabilities \citep{Richtmyer60, Meshkov69}. 

Our multidimensional conclusion are based mostly upon 2D Cartesian simulations as they are much computationally cheaper while the mechanisms of the dense shell distortion and destruction are naturally the same as in full 3D.
However, one should always keep in mind, that although the qualitative explanation (namely, which kind of instability arises, by which kind of perturbation it is excited and what to do to avoid its undesirable artificial
triggering) derived from 2D-calculations remains correct, its quantitative predictions are likely not, just because real instabilities do not support any kind of symmetry either planar or axial. It is also obvious, 
that e.g. fragmentation depends strongly on the problem dimensionality. Therefore, to investigate realistic spectra of the shell fragments or fluctuations, full 3D-simulations with realistic 3D-perturbations are 
inevitably required. 

Nevertheless, this and analogous forthcoming papers seem to be useful for those researchers who are interested in global RCS dynamics or fragmentation mechanisms
in circumstellar media and who intensively use numerical simulations in their studies, regarding either supernovae or their remnants.

In the rest of the paper we will review briefly the main physical effects which are essential for RCS dynamics, physical and numerical stability
in section~\ref{sec:problem}, describe our numerical techniques in section~\ref{sec:num}, present results of 1D and 2D
simulations in section~\ref{sec:res}, followed by discussions (section~\ref{sec:disc}) and conclusions (section~\ref{sec:conc}).
Some additional information can be found in Appendices, high-resolution images and videos can be accessed via our group web-page  \url{http://dau.itep.ru/sn/radshocksnr/}.

\section{The problem statement}\label{sec:problem}

\subsection{Physical properties and stability of the RCS}\label{sec:physstab}

As a background for instabilities we consider either spherical or cylindrical RCS of a strong central explosion in a static uniform medium. 
Volumetric losses are treated as a sink term in the equation of energy: 
\begin{equation}
   \partial_t(\rho e)+\partial_i(\rho v_i e)+p\partial_i v_i=
  -\Lambda(T) n_\text{e}n_\text{H} \label{eq:rho_e_cool} 
\end{equation}
where $p$, $e$, $\rho$, $n_\text{e}$, $n_\text{H}$, are pressure, specific internal energy per unit mass, mass density, number densities of electrons and hydrogen, respectively, 
and $\Lambda(T)$ is a cooling function (CF) accounting for a total loss rate of plasma in a given state (temperature $T$, and chemical composition) normalized by 
$n_\text{e}n_\text{H}$\footnote{Actually, different authors may use different definitions of $\Lambda(T)$ normalized by different number densities: 
$n_\text{H}^2$ (which may appear better for weak ionization with $(n_\text{e}/n_\text{H})$ accounted in $\Lambda$), $(\rho/m_\text{p})^2$ where $m_\text{p}$ is the proton mass, $n_\text{tot}^2$ 
-- total particle density. Therefore, one should be careful about the proper normalization of the cooling function.}. 

The medium is represented as an ideal gas with constant adiabatic index ($\gamma=5/3$ or 1.4). Its molecular weight is taken $\mu=0.5$ 
for the sake of simplicity of the equation of state. Several different cooling functions $\Lambda$ corresponding to different compositions 
are tested.

The radiative cooling introduces a specific characteristic time-scale:
\begin{equation}\label{eq:cooltime}
  t_{\text{c}}=\frac{\rho e}{\Lambda(T) n_{\text{e}}n_{\text{H}}}=\frac{kT}{\mu n_{\text{e}}\Lambda(T)}\approx 2.2\times10^3\frac{T_{[5]}}{\eta_{[4]}n_{0}\Lambda_{[-22]}(T)}\text{~yr}, 
\end{equation} 
where in the last equality $\eta_{[4]}=\rho/(4\rho_0)$ is a post-shock gas compression in units of 4 ($\eta_{[4]}=1$ for strong adiabatic shocks with $\gamma=5/3$), 
$n_{\text{e}}=n_{\text{H}}$ and $T_{[5]}$ -- temperature in $10^5$~K, 
$n_{0}$ -- ambient number density of protons in cm$^{-3}$ and $\Lambda_{[-22]}$ -- the cooling function in $10^{-22}$ erg~s$^{-1}$~cm$^3$.

Initially, for the first several thousand years, 
the shocked fluid is very hot ($T_{[5]}\gg1$) and its radiative cooling is not very efficient ($\Lambda_{[-22]}\ll1$), 
so the cooling time $t_{\text{c}}$ is large compared with a dynamical 
time $t_{\text{d}}= R_\text{s}/D_\text{s}$ ($R_\text{s}$ denotes the shock radius and $D_\text{s}$ -- its speed) and an adiabatic cooling time:
\begin{equation}
t_{\text{ad}}\equiv-\frac{T}{\dot{T}}=-\frac{1}{\gamma-1}\frac{\rho}{\dot{\rho}},
\end{equation}
where density $\rho$, temperature $T$ and their time derivatives are considered for a given fluid element. Therefore, the flow is nearly adiabatic and self-similar with 
$R_\text{s}\propto t^{2/5}$ \citep[in spherical symmetry,][]{Sedov}, $t_\text{d}=(5/2)t$, 
\begin{equation}
  t_\text{ad}=\frac{1}{\gamma-1}\frac{R_\text{s}}{3}\frac{\gamma+1}{2D_\text{s}}=\frac{5}{3}t
\end{equation} 
for $\gamma=5/3$ in the linear velocity approximation $u\propto r$, and the post-shock temperature
\begin{equation}
  T\approx6.59\times10^7\left(\frac{E_{[51]}}{n_0t^3_{[\text{kyr}]}}\right)^{2/5}\text{~K},
\end{equation}
where $E_{[51]}$ is the explosion energy in $10^{51}$~erg, $\rho_0$ -- ambient density in proton mass per cm$^{-3}$, $t_\text{[kyr]}$ -- time in thousand years.

However, as at high temperatures $\Lambda(T)$ depends on $T$ weaker than $T$, the cooling time $t_{\text{c}}$ should shorten as the temperature decreases. 
When the post-shock cooling time $t_{\text{c}}$ becomes comparable to $t_{\text{ad}}$ neither adiabaticity nor similarity takes place, but the shock flow transits gradually into an
opposite limit $t_{\text{c}}\ll t_{\text{d}}$ which is again self-similar, though not adiabatic (see below). 

A key feature common to all realistic $\Lambda(T)$ profiles (see section~\ref{sec:num} below for an illustration) is a strong (and multipeaked) bump in the range of $10^4-10^7$~K clearly prominent 
against a regular free--free power law background ($\Lambda\propto\sqrt{T}$). On its descending right slope $\Lambda(T)$ drops quickly which leads to strong positive
dependence of $t_{\text{c}}(T)$ and results in a so-called thermal instability even in static interstellar medium~\citep{ThermInst}: the colder and denser matter is, the faster it loses the energy
and contracts more and more. 

The cooling will slow down either if $T$ drops well below the region of short $t_\text{c}$ (at the ascending slope of the bump, $10^4-10^5$~K,
$\Lambda(T)$ grows much faster than $T$, so as $T$ goes down $t_{\text{c}}$ increases rapidly and in some cases is able to reach $t_\text{c}>t_\text{ad}$ again).
In the SNRs a nearly isothermal cold dense shell should form \citep{Cox1972,Falle1975,Falle1981} because the post-shock temperature growth is constrained from above by growing losses 
(or shortening $t_\text{c}$) and therefore to increase the density is the only way to withstand the ram pressure of upstream matter. 

Another (but somewhat speculative) possibility to stop the catastrophic cooling is separation of the shell into small-sized dense and optically thick clumps so that the losses become 
non-volumetric there (the multidimensional fragmentation is essential, because the pure radial contraction of the shell does not increase its optical thickness). 

Heating by an ionizing photon background, superthermal 
particles, heat transfer from a hot inner gas, cosmic rays etc. should also play a certain role in setting a temperature lower limit and the compression stabilization, but we do not take it 
into account in present paper as these relatively slow processes can become important considerably later after the catastrophic cooling period.

At a sufficiently large distance from the centre of explosion, the expanding spherical isothermal thin shell can be described by 
a self-similar solution  \citep[see][for derivation and consideration of applicability of the similarity theory to the radiative 
blastwaves]{OstrikerMcKee88}. For the uniform medium of density $\rho_0$, adiabatic index $\gamma$ and the initial explosion energy $E_0$ it reads: 
\begin{equation}
  R_{\text{s}}(t)=R_{\text{c}}\left(\frac{\xi E_0}{\rho_0 R_{\text{c}}^5}\right)^{\eta/2}t^\eta ,
  \label{RtPDS}
\end{equation}
where $R_{\text{c}}$ is an arbitrary characteristic radius \citep[e.g., according to][`at which half the energy of the initial blastwave has been
radiated away']{OstrikerMcKee88} and dimensionless coefficients $\eta$ and $\xi$ define two branches of the solution:
\begin{equation}
  \eta=\frac{2}{2+3\gamma}, \quad \xi=\frac{3(\gamma-1)(2+3\gamma)^2}{8\pi}
  \label{etaPDS}
\end{equation}
for a `pressure driven snowplough' (PDS) when the pressure of hot rarefied interiors pushes the dense shell forward, and
\begin{equation}
  \eta=\frac{1}{4}, \quad \xi=\frac{24}{\pi}
  \label{etaOort}
\end{equation}
for a `momentum conserving snowplough' (MCS) when the interiors are cold and the shell moves just under 
its inertia. This is known as `Oort stage' of SNR evolution \citep{Oort1946,Petruk2006,Hnatyk2007}.

While the MCS can realize only at the very late stages of SNRs, the PDS seems indeed relevant to our problem, and for $\gamma=5/3$ it gives
widely mentioned law $R_{\text{s}}\propto t^{2/7}$ \citep{McKeeOstriker1977} and the transition to this 
solution was obtained by \citet{BlinnikovImUt1982}.
We have tested our 1D simulations against this law and obtained a good agreement (see subsection~\ref{sec:1D} below).

In spherical symmetry the shell may become extremely thin and dense with density ratio $\rho/\rho_0$ as high as several hundreds observed in simulations
on Lagrangian meshes \citep[see e.g.][]{Straka}. However, such high compression of spherical shells is unlikely to be reached in real 3D flows because of various instabilities. 

If being accelerated by the inner pressure (this can take place during re-establishing of a hydrodynamical balance after the catastrophic cooling onset), 
the shell interface (separating hot gas with $t_{\text{c}}\gg t_{\text{d}}$ and the cool one with $t_{\text{c}}<t_{\text{d}}$) should be Rayleigh--Taylor unstable 
\citep{BernsteinBook78}. 

While the shock velocity $D_\text{s}$ decreases from the adiabatic dependence $D_{\text{s,a}}\propto t^{-3/5}$ to the PDS one with cooling $D_{\text{s,c}}\propto t^{-5/7}$, 
the velocity of matter can eventually grow from $2D_{\text{s,a}}/(\gamma+1)$ to nearly $D_{\text{s,c}}$ (if the transition is rapid enough and the latter can be still greater than the former). 

\citet{Falle1981} has shown that after the cooling onset the RCS is overstable for radial pulsations. It is also shown (see fig.~5 of that work) that, during the shell formation,
several secondary shocks arise behind the forward one. They can collide with the shell and provide additional impulses of acceleration to it, 
which suggests that the Richtmyer--Meshkov instability can take place there.

For stationary shocks in uniform media and non-spherical symmetric perturbations, the question of stability of highly compressed thin 
shell (bounded by shocks either from one or from both sides) 
against rippling deformations was addressed in analytical works of Vishniac in linear \citep{Vishniac1983, 
VishniacRyu89} and non-linear \citep{Vishniac1994} formulations.

The problems were considered in a so-called small bending angle limit, when either an amplitude of 
bends assumed to be much less than the shell's scale of height $L_{\text{s}}$, or their wavelengths should be much larger than $L_{\text{s}}$.
Additionally, governing equations were
formulated in terms of column densities of mass, momentum and shear 
(or tangential velocity gradient, only in the non-linear case) in order to avoid detailed treatment of a 
shell vertical (radial) structure. 

The PDS- and MCS-shells were shown to be linearly overstable, i.e. liable to progressing column density oscillations caused by tangential mass transfer 
along oblique parts of the corrugated shock surface from advancing arcs towards trailing ones. Depending on the shock Mach number~\citep{VishniacRyu89}, the instability growth
was obtained for spherical-harmonic numbers of perturbations from several units to several hundreds, peaking around several tens.

Non-linear instabilities were discovered in two-shock bounded stationary slabs (a non-linear thin shell instability, NTSI) and in decelerating MCS-shells (a non-linear
deceleration instability, NDI). They were explained by a non-linear feedback of shear motions transporting mass and momentum within the shell on to 
its bending and breathing modes. 

The NTSI results in gradual growth of bending amplitude, while the NDI somewhat resembles the linear overstability, though with non-linear resonant coupling 
of bending and breathing modes and excitation of their harmonics. Due to the efficient harmonic excitation, even for relatively long-scale initial 
perturbations the NDI can produce a complex structure inside the shell, where high-density spots tend to overtake the shell mid-surface because of their inertia 
(in contrast to the linear overstability where matter is piled up in the trailing ripples).

From this variety of the thin shell instabilities, the PDS overstability (hereafter -- PDTSO for the pressure driven thin shell overstability) 
is the only one which is completely relevant to our problem. 
Additionally, as we deal with the two-shocks bounded shells, the NTSI also may be of particular interest, though with the difference that our shell is not stationary.

In addition to these hydrodynamic instabilities from the analytical works of Vishniac, we should take into account the thermal instability. 
If it acts, the denser regions of the shell will lose their energy quicker than the rest ones, and thus the restoring force should be somewhat suppressed. 
Obviously, this may lead to clumping acceleration and successive fragmentation of the flow on shorter time-scales than expected from the pure-hydrodynamical 
treatment \citep[however, as it is pointed out by][, under certain circumstances volumetric losses may not have enough time to modify 
instability growth considerably]{KrasnobaevTagirova}. 

Stability of radiative thin shells of the SNRs was studied by \citet{Blondin98} with high resolution two-dimensional simulations (however, constrained to a narrow sector of a polar grid, so 
only short-wavelength perturbations were available). Those authors interpreted observed early (after the shell formation) oscillations as PDTSO \citep[however, with a rate inconsistent with 
predictions of the linear-theory dispersion relation of][]{VishniacRyu89}, and later monotonously growing ripples as NTSI of the shell bounded by forward and reverse shocks. 

Based on global 2D simulations and careful inspection of 1D-ones, we are going to show that this explanation is not comprehensive, and other physical mechanisms may play a more important role
in the perturbations amplification, especially if the shell structure is resolved and taken into account.

\subsection{Perturbations affecting numerical solution}\label{sec:numstab}

We perform non-linear global numerical simulations. Numerical solutions
are inevitably affected by computational imperfections of calculations (e.g. finite mesh resolution or some systematic errors of forward-integration of differential 
equations) which can be effectively considered as perturbations of an \textit{a priori} unknown (or determined only roughly) shape and strength, because the 
numerical perturbations actually arise from the solution itself during the time it is being obtained. 
This means that the simulated flow reacts not to initial deviations from a stable state like in analytical setups but to a complex continuous transient perturbing force. . 

The real flow is smooth and continuous (even at the shock jump tangential continuity should hold), while the discrete one is
treated as a superposition of separate piecewise-continuous `elementary flows'  along 
orthogonal directions (in Eulerian finite-volume schemes on Cartesian grids) between neighbouring rectangular cells. Numerical noise inherent to this representation acts 
like a short-wavelength bending perturbation for the thin shell, a relative strength of which depends on 
how fine the mesh can resolve the characteristic thickness of the real flow.

In widely used Godunov schemes, such a spatial decomposition results in that the Riemann problem in most cases is solved only along basic directions.
This leads to certain deviations of evaluated paths and speeds of hydrodynamical information propagation from apparent characteristics depending 
on whether the flow is aligned to the grid lines or not, effectively making the schemes not exactly upwind. This can also bend the modelled curved 
shells in global simulations on rectangular meshes.

Some flows are stable against such  perturbations \citep[like strong adiabatic blastwaves in uniform media with $\gamma$ well above 1.2, 
stability of which is proven analytically, experimentally and numerically, see e.g.][]{VishniacRyu89}, and others may be not. 
Taking into account complex physics of the RCS shells,
they are likely to be of the latter type, and probably their stability against the most important numerical perturbations observed in simulations can not be studied 
analytically \citep[i.e. like for example `carbuncle' or `odd-even' instabilities of pure-hydrodynamic calculations discussed in]
[and similar papers]{GodunovInstab, LiouStab, LiskaWendroff}.  

That is why special numerical experiments are required to assess susceptibility of the simulated RCSs to the instabilities contained both in physical equations and 
in their numerical realisations.

\section{Calculational techniques}\label{sec:num}

In order to collect the most general and universal information from numerical simulations of RCSs, 
we have conducted our experiments using several different codes, different grid types,
different kinds of symmetry and different cooling functions.

The codes, described below, are independent and solve hydrodynamic equations explicitly in Eulerian formulation in rectangular, polar and spherical coordinates in planar, 
axial or spherical symmetry with the radiative cooling treated implicitly. Two of them are able to apply adaptive mesh refinement (AMR) algorithms. 

\begin{enumerate}
\item The {\small FRONT3D} code\footnote{http://dau.itep.ru/sn/front3d} is a
  general purpose parallel astrophysical code developed in our group. It
  can take into account various physical processes of ideal hydrodynamics, magnetohydrodynamics,
  thermoconductivity, nuclear burning, etc. For hydrodynamical problems
  it utilizes a finite-volume approach with a MUSCL-scheme and several realisations of Riemann-solvers. In present simulations the HLLC-solver is applied without any H-correction \citep[unlike][]{KimOstriker14}.

  A physical process splitting technique is used for complex problems. For purposes of the RCS modelling, each
  time-step is subdivided into independent sub-steps of adiabatic advection and local radiative cooling of static matter. 
  The cooling is treated in an exact implicit manner suggested by~\citet{Townsend_ApJS_2009} and described in Appendix~\ref{app:CoolTown}.
  
\item The {\small PLUTO} code is a well-known public open-source code for
  astrophysical community \citep{Pluto}. 
  In our tests we used it only in Cartesian geometry with HLL or two-shock Riemann-solver, and RK3 time-stepping.
  
\item The {\small FLASH} code is another well-known astrophysical code~\citep{FLASH2000}. We used the directionally
split PPM (piecewise-parabolic method) solver on adaptively refined meshes both in Cartesian and polar geometries.
For interface flux calculations we employed a regular (two-shock Riemann) solver and a hybrid Riemann-solver which uses the
regular solver in most regions but switches to an HLLE within shocks.
\end{enumerate}

As for grids and symmetry, we performed runs mostly in 2D on Cartesian and polar meshes, with a few 3D tests 
with no principal distinctions (but rather expensive),
and 1D calculations also carried out as reference ones for 2D results interpretation. 

2D-models were set up either in axial or in planar symmetry, so that a circle in them represented a mid-plane cross-section of a sphere or an infinite cylinder 
respectively. 

Axisymmetric models used either an `$RZ$' Cartesian grid based upon the axis of symmetry ($Z$-dimension) and its perpendicular ($R$-dimension), or an `$R\theta$' (or `Rth') 
polar grid parametrized by an inclined radius $R$ and a polar angle $\theta$ with respect to the polar axis. 

Planar-symmetric models were set up in simple rectangular `$XY$' coordinates or polar `$R\varphi$' 
(`Rph', radius vs. azimuthal angle) ones. Computation domains in all cases represented a quadrant with its grid origin situated in a centre of explosion. 

1D simulations assumed either a spherical symmetry for comparison with $RZ$- and $R\theta$-runs, or a cylindrical one -- for $XY$ and $R\varphi$. 

The $XY$- and $R\varphi$-calculations have the advantage that they are carried out in the transverse plane and there is no dedicated direction 
(like the polar axis in $RZ$ or $R\theta$). Therefore, if there were any symmetry in perturbations, it should be better reproduced in these setups. 
Additionally, $XY$-calculations better demonstrate whether one or another numerical scheme is balanced with respect to basic directions or not. 

Indeed, our $RZ$-runs did show a considerable asymmetry of the instability pattern with respect to grid basic directions (see below), and the $XY$-ones did not.
That is why, in order to facilitate interpreting of results, the most of our simulations were performed in $XY$ and $R\varphi$, despite the fact that an 
`infinite cylinder explosion' seems somewhat unphysical. On the other hand, we are going to study quite general features of RCS evolution like an adiabatic-to-radiative shock transition under 
non-stationary cooling, the thin shell formation and possible instabilities, which should be common for RCSs in different symmetries.

$XY$-calculations can also be considered analogous to what one should expect from 3D-Cartesian simulations (in terms of instability development): both use rectangular shaped elements and possess equal 
translation symmetry (due to absence of dedicated points or axes in equations). That is why equatorial plane slices of the 3D-explosion on the Cartesian grid presented by \citet{KimOstriker14} look much 
more similar (in terms of symmetry of bends, see below) to our $XY$-cylindrical RCS simulations than to spherical $RZ$-ones. 

The initial explosion was modelled by a `thermal bomb', i.e. deposition of $E_0$ in a form of internal energy of matter of the unperturbed density $\rho_0$ within a certain radius 
$R_0$ from the grid origin.

Several different cooling functions (CF) were tested.
The first one was a semi-analytic pure hydrogen CF (for $\gamma=5/3$, $\mu=0.5$)  adopted from \citet{Straka} (hereafter Straka-CF):
\begin{equation}
 \Lambda_{\text{S}}= A\times 10^{-2\left[\log_{10}(T/T_0)\right]^2}+BT^{1/2}.
\label{Straka}
\end{equation}
where $T_0=1.93\times 10^5$~K, $A=9.982\times 10^{-22}$ and $B=6.973\times10^{-28}$ in cgs units. 

\begin{figure}
  \centering
  \includegraphics[width=0.95\linewidth]{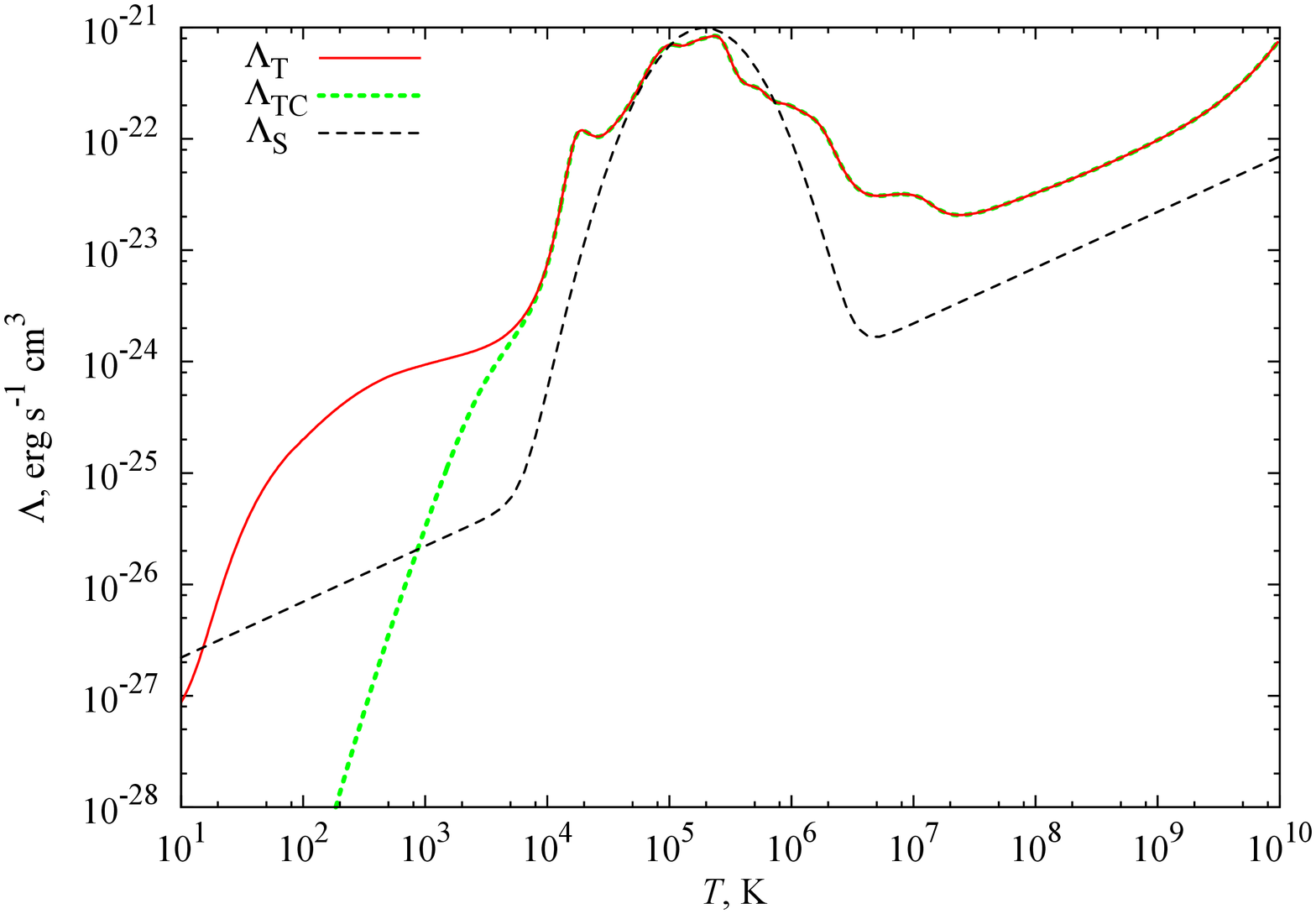}
  \caption{\label{fig:CF} Comparative plot of three used cooling functions: the Table-CF (red solid), Cut Table-CF (green dotted) and the Straka-CF (black dashed).}
\end{figure}

The second one was taken from the {\small PLUTO4.1} code (a tabulated cooling option) and 
corresponds to solar metallicity 
(hereafter Table-CF, or $\Lambda_\text{T}$). As the lowest temperature available in the Table-CF is 10~K, we used this value as a low temperature limit
for cooling iterations.  

As we do not take into account recombination directly, assuming instead $n_\text{e}=n_\text{H}$ whatever the temperature, 
thus we need to ``suppress'' the table cooling function at low temperatures in order to obtain more realistic losses. 
It seems improper to limit the cooling with the warm ISM temperature (e.g. $\approx10^4$~K), because in fact the equilibrium temperature results from balance between several processes of radiative cooling and heating 
(radiative, conductive, cosmic rays, MHD-dissipation -- and still real circumburst medium may appear out of equilibrium) acting on different time-scales,
efficiencies of which depend on density in different ways. The limiting temperature for the dense shell should be lower than for the rarefied ISM. 
According to the ISM equation of state \citep[][]{Wolfire}, its equilibrium temperature for $\log_{10}(n)=2.5$ is around 30~K.

In the case of the RCS thin shell, the heating-cooling balance, obviously, is far from stationarity
and equilibrium (al least at the early stages of the dense shell formation the fast cooling should dominate), so it may easily overcool 2-3 orders of magnitude below $10^4$~K. 
Therefore, when studied the thin shell structure at high resolution (i.e. when very high densities were expected),
we introduced a cut-off Table-CF (hereafter Cut Table-CF, or $\Lambda_{TC}$), i.e. multiplied by a smooth factor converging to 1 for temperatures above several $10^3$~K and becoming $\ll1$ below 
$10^3$~K: 

\begin{equation}
  \Lambda_\text{TC}= \frac{1}{2} \left[1+\tanh\left(\frac{\log_{10}(T_{[K]})-A}{B}\right)\right]\Lambda_\text{T},
\end{equation}
with A=3.5, B=0.3. These three cooling functions are plotted in Fig.~\ref{fig:CF}.

\begin{table*}
\centering
\caption{\label{tab:models} List of models
}
\begin{tabular}{l|c|c|c|c|c|c|c|c}
 \hline
 Mod. name   & Grid$^a$           & $R_{\max}\!$ $^b$ & $R_0\!$ $^c$ & $E_0\!$ $^d$ & $\rho_0\!$ $^e$ & $\gamma$ & $\Lambda$ & Code \\
             &                    & pc  & pc     & foe or foe~pc$^{-1}$  & $m_{\text{p}}$ cm$^{-3}$
                                                                          &     &          &                  \\ 
 \hline
 E1S-S400   & 1D spherical 400   & 50  & 2      & 1.28              & 1  & 5/3 & Straka   & \small{FRONT3D}  \\
 E1T-S1600 & 1D spherical 1600  & 100 & 2      & 1.28              & 1  & 5/3 & Table    & \small{FRONT3D}  \\
 E1S-RZ1600 &  $RZ$ 1600$^2$     & 50  & 2      & 1.28              & 1  & 5/3 & Straka   & \small{FRONT3D}  \\
 E1T-RZ9  &  $RZ$ 512$^2$      & 50  & 2      & 1.28              & 1  & 5/3 & Table    & \small{FRONT3D}  \\
 E2T-XY9  &  $XY$ 512$^2$      & 80  & 2      & 0.48L             & 1  & 5/3 & Table    & \small{FRONT3D}  \\
 E3T-C18  & 1D cylindrical $2^{18}$  & 50  & 2      & 0.087L             & 1  & 5/3 & Table    & \small{FRONT3D}  \\
 E3TC-C18  & 1D cylindrical $2^{18}$ & 50  & 2      & 0.087L             & 1  & 5/3 & Cut Table    & \small{FRONT3D}  \\
 E3T-C19A & 1D cylindrical A$2^{19}$  & 50  & 2      & 0.087L             & 1  & 5/3 & Table   & \small{FLASH4.2}  \\
 E3TC-C19A & 1D cylindrical A$2^{19}$ & 50  & 2      & 0.087L             & 1  & 5/3 & Table   & \small{FLASH4.2}  \\
 E3TC-C19AP$^f$ & 1D cylindrical A$2^{19}$ & 50  & 0.002      & 0.087L             & 1  & 5/3 & Table   & \small{FLASH4.2}  \\
 E3S-XY9  &  $XY$ 512$^2$      & 80  & 2      & 0.087L             & 1  & 5/3 & Straka   & \small{FRONT3D}  \\
 E3T-XY9  &  $XY$ 512$^2$      & 80  & 2      & 0.087L             & 1  & 5/3 & Table    & \small{FRONT3D}  \\             
 E3TC-XY11  &  $XY$ 2048$^2$      & 50  & 2      & 0.087L             & 1  & 5/3 & Cut Table    & \small{FRONT3D}  \\
 E3T-Rph8
             &  $R\varphi$ $256_R\times128_\varphi$  
                                  & 80  & 2      & 0.087L            & 1  & 5/3 & Table    & \small{FRONT3D}  \\
 E4S-XY8  &  $XY$ 256$^2$      & 10  & 0.0137 & $3\times10^{-5}$L & 1  & 5/3 & Straka   & \small{PLUTO4.1}   \\
 E4F-XY8  &  $XY$ 256$^2$      & 10 & 0.0137  & $3\times10^{-5}$L & 1  & 5/3 & Free--free & \small{PLUTO4.1}   \\
 E4T-XY10 &  $XY$ 1024$^2$     & 5   & 0.0137 & $3\times10^{-5}$L & 1  & 5/3 & Table    & \small{FRONT3D} \\
 E4S-XY9A   &  $XY$ A512$^2$   & 5   & 0.137 & $3\times10^{-5}$L & 1  & 5/3 & Straka   & \small{FLASH4.2} \\
 E4T-XY9A    &  $XY$ A512$^2$         & 5   & 0.137 & $3\times10^{-5}$L & 1  & 5/3 & Table    & \small{FLASH4.2} \\
 E4T-XY12A   &  $XY$ A4096$^2$     & 3   & 0.137 & $3\times10^{-5}$L & 1  & 5/3 & Table    & \small{FLASH4.2} \\
 E4T-C12    &  1D cylindrical $2^{12}$    & 5   & 0.0137 & $3\times10^{-5}$L & 1  & 5/3 & Table    & \small{FRONT3D} \\
 E4S-Rph8A   &  $R\varphi$ A$256_R\times256_\varphi$
                                & 5   & 0.137 & $3\times10^{-5}$L & 1  & 5/3 & Straka   & \small{FLASH4.2} \\
 E4T-Rph8A   &  $R\varphi$ A$256_R\times256_\varphi$   & 5   & 0.137 & $3\times10^{-5}$L & 1  & 5/3 & Table    & \small{FLASH4.2} \\
 E4T-Rph12AD$^g$ &  $R\varphi$ A$2^{12}_R\times2^{12}_\varphi$  & 3   & 0.137 & $3\times10^{-5}$L & 1 & 5/3 & Table    & \small{FLASH4.2} \\
 E5T-XY9A    &  $XY$ A512$^2$         & 3   & 0.137 & $5\times10^{-5}$L & 10 & 1.4 & Table    & \small{FLASH4.2} \\
 E5T-Rph8A   &  $R\varphi$ A$256_R\times256_\varphi$  & 3   & 0.137 & $5\times10^{-5}$L & 10 & 1.4 & Table    & \small{FLASH4.2} \\
 E6F-XY9  &  $XY$ 512$^2$      & 50 & 0.684  & $1.3\times10^{-3}$L & 1 & 5/3 & Free--free & \small{PLUTO4.1}   \\
 \hline
 \multicolumn{9}{l}{$^a$Dimensionality, symmetry and apparent or effective (in the cases of AMR usage labelled by A) number of grid cells} \\
 \multicolumn{9}{l}{$^b$Linear grid size} \\
 \multicolumn{9}{l}{$^c$Radius of the initial energy release domain. } \\
 \multicolumn{9}{l}{$^d$Explosion energy for spherical explosions or energy per unit length for cylindrical ones (C, XY, Rph),} \\
 \multicolumn{9}{l}{~~1 foe = $10^{51}$~erg, and foe~pc$^{-1}$ is foe per parsec, the latter specific energy values are labelled by L in the table} \\
 \multicolumn{9}{l}{$^e$Ambient density in proton mass per cubic centimetre} \\
 \multicolumn{9}{l}{$^f$P is for a point-like explosion, since its initial domain is much smaller that in other models} \\
 \multicolumn{9}{l}{$^g$Density perturbations were placed before the RCS.} \\
\end{tabular}
\end{table*}

We  also used a simple \verb|POWER_LAW| option in several simulation with {\small PLUTO} (see below), which is just a free--free ($\sqrt{T}$) cooling law.

Finally, initial conditions differed in the burst energy $E_0$, ambient matter density $\rho_0$ and its adiabatic index $\gamma$.

We summarize the properties of our models discussed below in Table~\ref{tab:models}. Each model name consists of a part of physical properties 
($E_0$, $\rho_0$, $\gamma$ combined into E1, E2, etc. code names of explosions and a CF type: S, T, TC or F) and a part of calculational setup (dimensions, 
grid kinds, numbers of cells and some-times special comments). Number of cells (per axis) is code-named either with its apparent value or with its binary logarithm (if integer), 
for the AMR-runs this is not a real number (which is variable) but a ratio of the overall grid span to the finest mesh cell width.

Generally, the E1-class represents a typical supernova. E2-models are derived from E1, however
with initial explosion domain changed from a sphere of radius $R_0$ to a cylinder of radius $R_0$ and length $L=4R_0/3$ (simply in order to hold the internal 
energy density the same). E3 are lower-energy and shorter-times versions of E2 (therefore they were preferred in calculations). 
Models E4--E5 are constructed to provide a much finer spatial resolution (with comparable number of cells), however, 
this costs a reduced energy yield. Additionally, E5 aimed to speed up the cooling and also to study possible effect of increased ambient pressure. 

Model E6 is an auxiliary one and was run just to test and illustrate the instability emergence even with monotonous $\Lambda(T)$ in certain environments.

\section{Numerical results}\label{sec:res}
 
\subsection{1D simulations}\label{sec:1D}

In order to interpret results of multidimensional calculations correctly, it is useful to understand an order of key events by the example of various 1D-simulations as they are computationally cheaper,
more flexible and their profiles are more suitable for the comprehensive qualitative and quantitative analysis. 

We performed series of 1D-runs in spherical and cylindrical setups. Generalization of their results can be
subdivided into three stages: kinematics of the shock (radii, velocities), global picture of the adiabatic-to-radiative cooling transition and high-resolution studies of the thin shell 
close environment.

\subsubsection{Kinematics of spherical RCS and effect of low-temperature cooling treatment}\label{sec:kinematics}

\begin{figure*}
  \includegraphics[width=1.\linewidth]{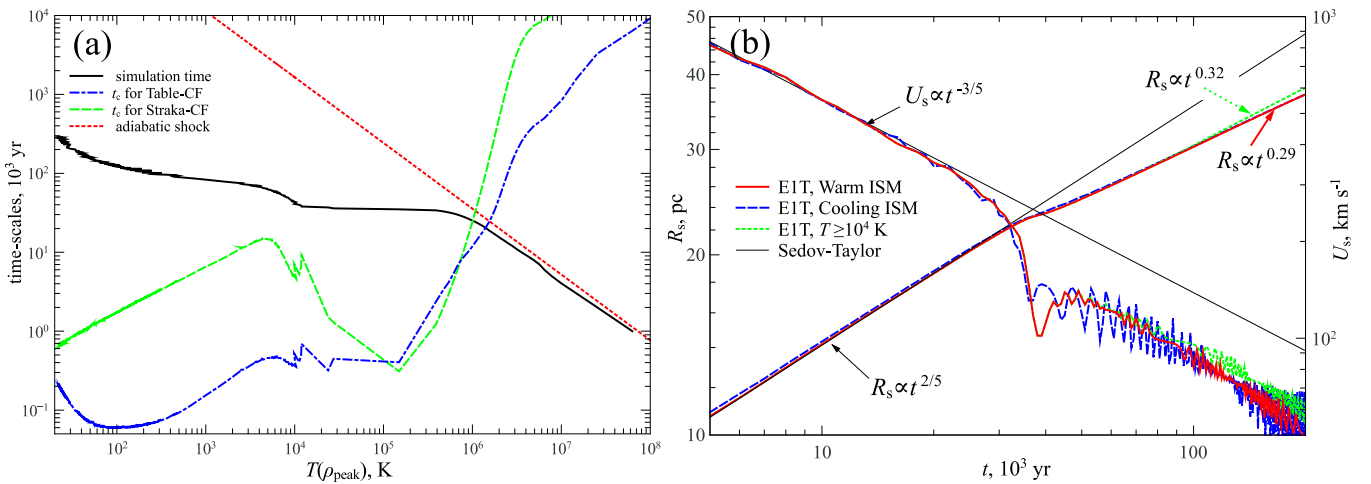}
  \caption{\label{fig:1D}Cooling and expansion dynamics of the E1T-S1600 spherical shock compared with analytical predictions. 
  (a) -- Temperature course at density maximum (solid black), self-similar Sedov solution for the same initial parameters (red dotted) and cooling time-scales $t_\text{c}$
  for relevant calculated temperatures and densities with the Table-CF (blue dash-dotted) and the Straka-CF (green dashed) for comparison.
    (b) -- Radii and velocities of the forward shock in different low-temperature cooling approaches: warm ISM (fixed $T=10^4$~K) plus unconstrained post-shock cooling (red solid), 
    unconstrained post-shock plus ISM cooling (blue dashed), warm ISM plus temperature limit $T\geq10^4$~K everywhere (green dotted). The Sedov solutions are also plotted as thin solid black lines.}
\end{figure*}

\begin{figure*}
   \centering
    \includegraphics[width=1.\linewidth]{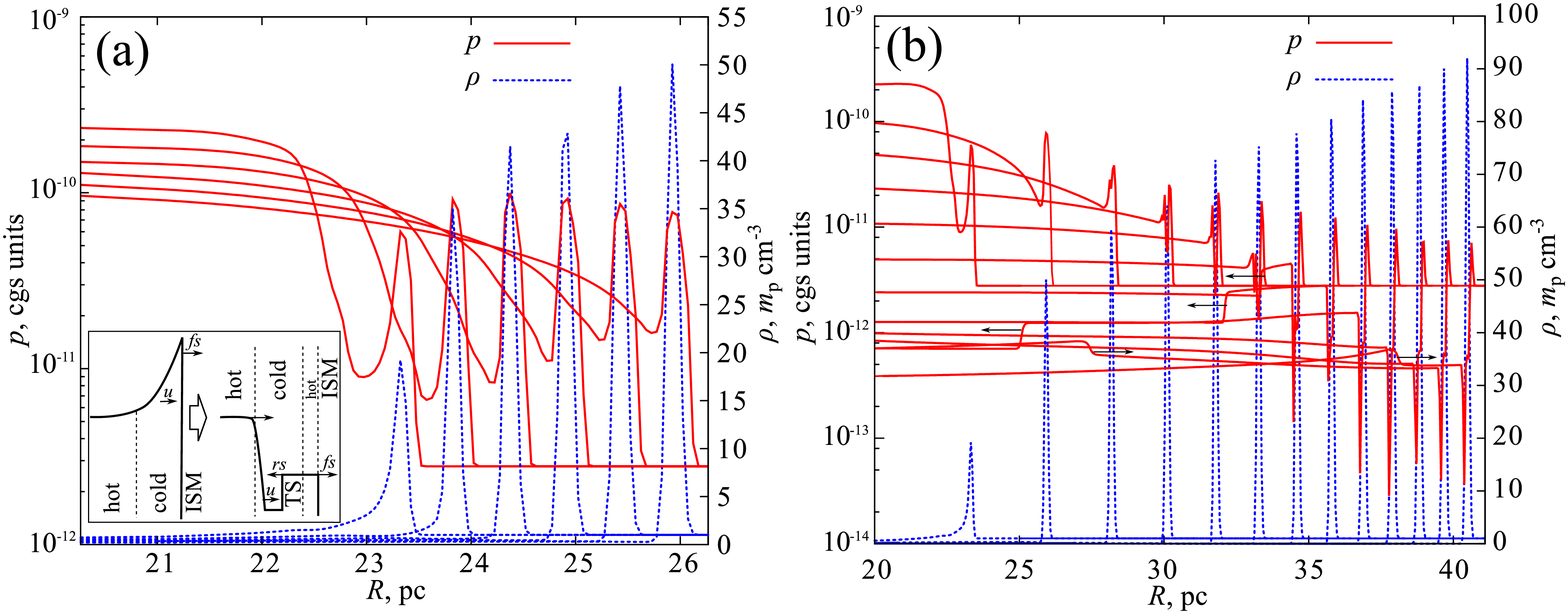}
  \caption{\label{fig:StructS} Pressure ($p$) and density ($\rho$) radial profiles around the thin shell of the E1T-S1600 run. (a) -- early stages (time-step is uniform) and a cartoon of pressure 
  drop development during the catastrophic cooling. 
  (b) -- complete picture since the beginning of the shell formation until the late stages (time-step is uniform, but larger than in (a)). 
  Thin black arrows depict the hot flow termination (reverse) shock moving inwards and then reflected.}
\end{figure*}

First of all, we have tested the results of our codes against spherical-symmetric analytical predictions. 

Fig.~\ref{fig:1D}a illustrates typical time-scales and temperature dynamics of the post-shock fluid (more exactly, at the maximum of density) in the spherical explosion E1T. 
We have plotted $t_\text{c}$ for both cooling functions  used in our modelling (compression degrees were taken from run E1T-S1600), 
as well as time vs. post-shock temperature dependence for a Sedov-Taylor adiabatic solution 
(for $E_0$ and $\rho_0$ of E1T model) and that of the real simulation. One can see, that equating $t_\text{c}=t_\text{ad}=(5/3)t$ provides a good estimation of the catastrophic cooling onset time.
 
 Another conclusion is that  for adopted cooling functions the fluid in the thin shell, once entered the catastrophic cooling regime, then never leaves it and permanently cools and contracts. 
 However, the Straka-CF provides a considerable increase of $t_\text{c}$, and in some circumstances (namely, lower densities or lower compressions, which are likely in the case of slower cooling rates
 of Straka-CF) may stabilize the collapse for a while (when $T$ is around several $10^3$~K). Therefore, we claim, that the low-temperature wing of $\Lambda(T)$ and 
 low-temperature cooling treatment are very important and may be crucial in reproducing of late-time dynamics of radiative SNRs.
 
 We can additionally illustrate this with the fact that \citet{Blondin98} and \citet{Falle1975, Falle1981} did not obtained the PDS asymptotics ($R_\text{s}\propto t^{2/7}$) in their calculations 
 having the time power index around $0.3-0.33$ 
 (the latter is the closest to our E1T model parameters), but \citet{Straka} did and we do as well. Moreover, at the latest stages, we and Straka see even `softer' shock dynamics converging to
 $R_\text{s}\propto t^{0.25}$ of the Oort solution. 
 
 \citet{Blondin98} explain this to be due to finite (non-zero) ambient temperature and pressure (unlike those previous simulations where these were negligible), 
 which yields higher post-shock temperatures and pressures and therefore pushes the shock front somewhat faster.  
 
 We had run two modifications of E1T-S1600 calculation, one with warm ($T_0=10^4$~K) and another with cold 
 ISM (i.e. it was allowed to cool from $10^4$~K unconstrainedly due to radiative losses), and found no significant difference. However, as soon as we had forbidden  
 the fluid to cool below $10^4$~K (by means of a special logical condition), we indeed reproduced the $R_\text{s}(t)$ of \citet{Blondin98}. Additionally, as fluid temperature reaches this limit, the shell
 stops contracting and even begins to expand, decreasing its peak density. So, very likely the `regain of elasticity' to be not due to the excessive post-shock entropy production, 
 but just from the low-temperature cut-off, and namely the fact that Falle and \citet{Blondin98} did not prolong $\Lambda(T)$ below $10^4$~K unlike Straka and us. 
 
 We will return to the question of cooling limiting below, and now refer the reader to Fig.~\ref{fig:1D}b where we show
 our obtained shock radii and estimated velocities (simple centred time-derivatives of the output tables $R_\text{s}(t)$) for different types of low-temperature cooling. 
 
 Fig.~\ref{fig:StructS}-\ref{fig:StructC} illustrate the many-shock structure of the flow of the forming thin shell and its later evolution in spherical 
 (E1T-S1600, Fig.~\ref{fig:StructS}) and high-resolution cylindrical (E3TC-C19A, E3TC-C19AP, Fig.~\ref{fig:StructC}) calculations.
 
 The sub-panel of Fig.~\ref{fig:StructS}a shows schematically how the thermal instability comes into action. A cooling wave `swallows' the post-shock regions quickly, but `sticks' when 
 reaches the hot interiors. The pressure drops quickly, but the fluid retains its inertia and drives the forward shock (FS or `fs' on the scheme) through the ISM. The ISM passes through the FS-jump
 much quicker compared to its cooling time (at relevant densities and temperatures) so the FS is nearly adiabatic. That is why the dense shell begins to form at the bottom of the pressure gap 
 a cooling length  $\lambda_\text{c}\sim (D_\text{s}-u_\text{ds})t_\text{c,s}$ behind the forward shock (see Fig.~\ref{fig:StructC}a). 
 
 Pressure at the forward shock eventually becomes lower than that of the hot rarefied interiors (not involved in catastrophic cooling), so the pressure gap is asymmetric and its rear negative gradient 
`feeds' the dense shell with the cooling fluid (and its momentum) at a much higher rate than a forward cooling region (FCR). This is indeed a `snowplough' from the slope of which matter slides 
down to its foot. Actually, this not only weakens deceleration of the cooling fluid but also can provide a positive acceleration to it for a while, making it potentially unstable.  

Once the developed dense shell appears able to sustain its width (either by the thermal or magnetic pressure grown up due to compression, or by means of numerical effects like finiteness of the cell size), 
then the reverse or termination shock should arise, propagating inwards through the light hot fluid. 
The latter is reheated again (adiabatically) passing the reverse shock and comprises a rear cooling region (RCR) before terminating at the dense shell.

 \subsubsection{Thermal instability: a multishell structure formation and collapse}\label{sec:collapse}
 
\begin{figure*}
  \centering
  \includegraphics[width=1.\linewidth]{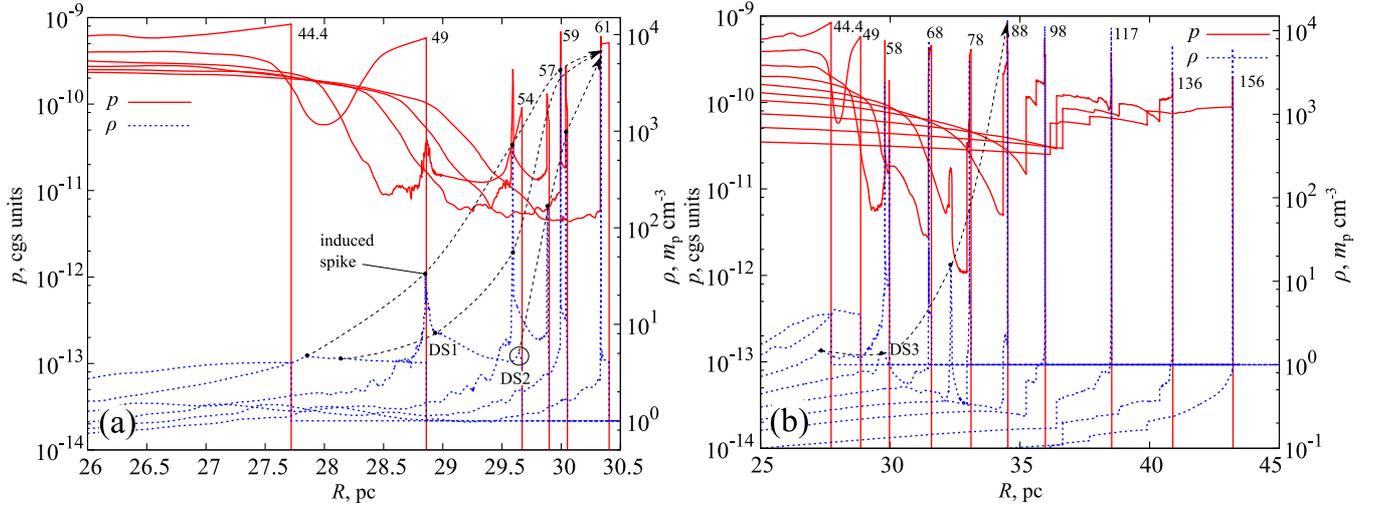}
  \caption{\label{fig:StructC}Pressure ($p$) and density ($\rho$) radial profiles of the E3TC-C19A flow forward part. (a) -- $p$ and $\rho$ during the shell 
  formation, numbers around the peaks indicate time in $10^3$~yr, dashed black arrow illustrates evolution of density peaks (labelled with DS1, DS2, DS3) and their contraction into the dense shell. 
  (b) -- the same but for longer times.}
\end{figure*}

\begin{figure*}
  \centering
  \includegraphics[width=1.\linewidth]{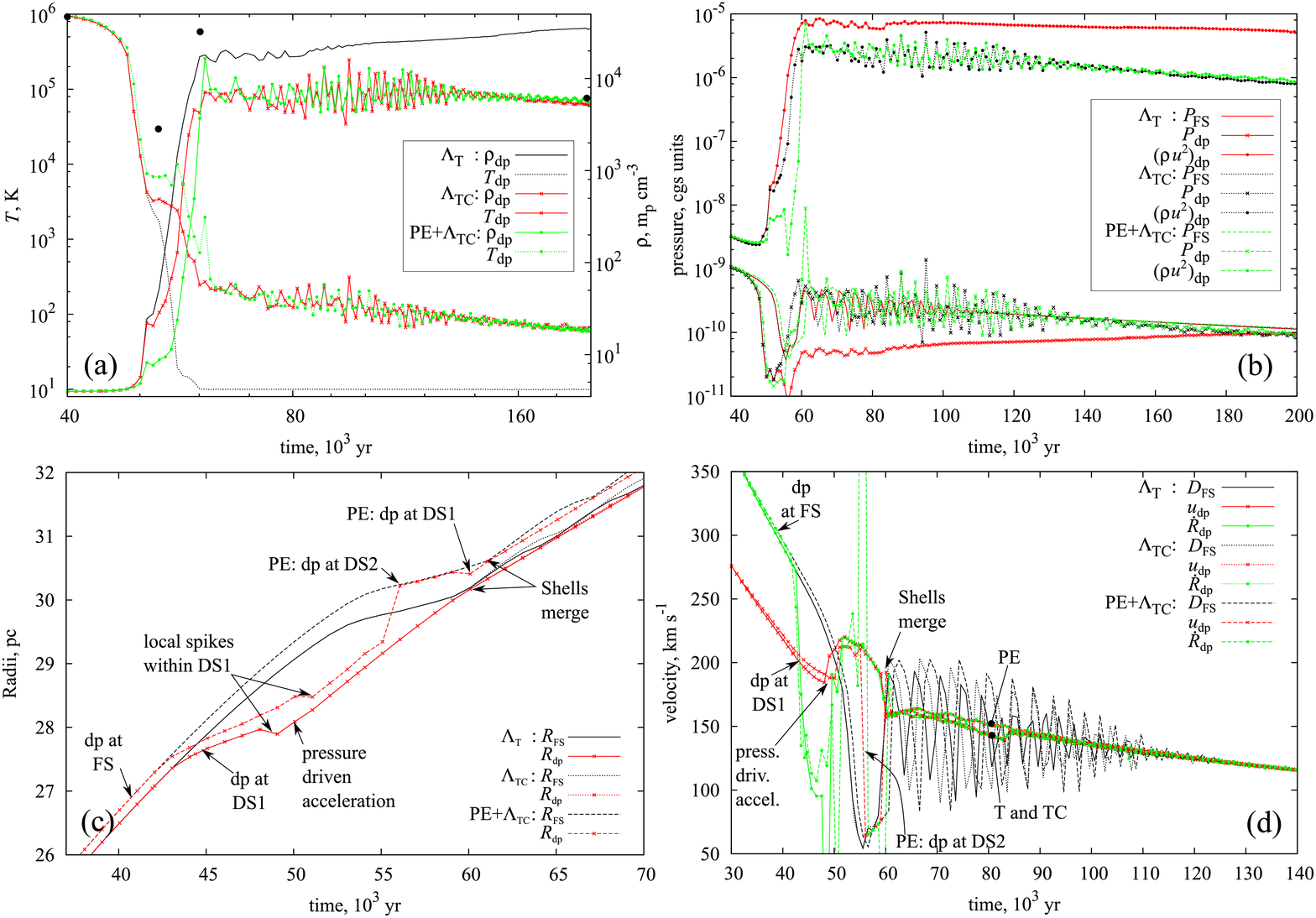}
  \caption{\label{fig:Diagnostics} Time-evolution of state and kinematics at the dense shells and forward shocks in different high-resolution E3-models: $\Lambda_\text{T}$ -- E3T-C19A, $\Lambda_\text{TC}$ -- E3TC-C19A, 
  $\text{PE}+\Lambda_\text{TC}$ -- E3TC-C19AP. (a) -- density $\rho_\text{dp}$ and temperature $T_\text{dp}$ during at the highest density peak.  (b) -- forward shock pressure $P_\text{FS}$, thermal pressure $P_\text{dp}$ and
  kinetic energy density $(\rho u^2)_\text{dp}$ at the highest density peak. (c) -- radii of the forward shock $R_\text{FS}$ and the highest density peak $R_\text{dp}$. 
  (d) -- forward shock speed $D_\text{FS}$, fluid velocity at the highest density peak $u_\text{dp}$ and speed of propagation of this peak $\dot{R}_\text{dp}$.}
\end{figure*}

\begin{figure}
  \centering
  \includegraphics[width=1.\linewidth]{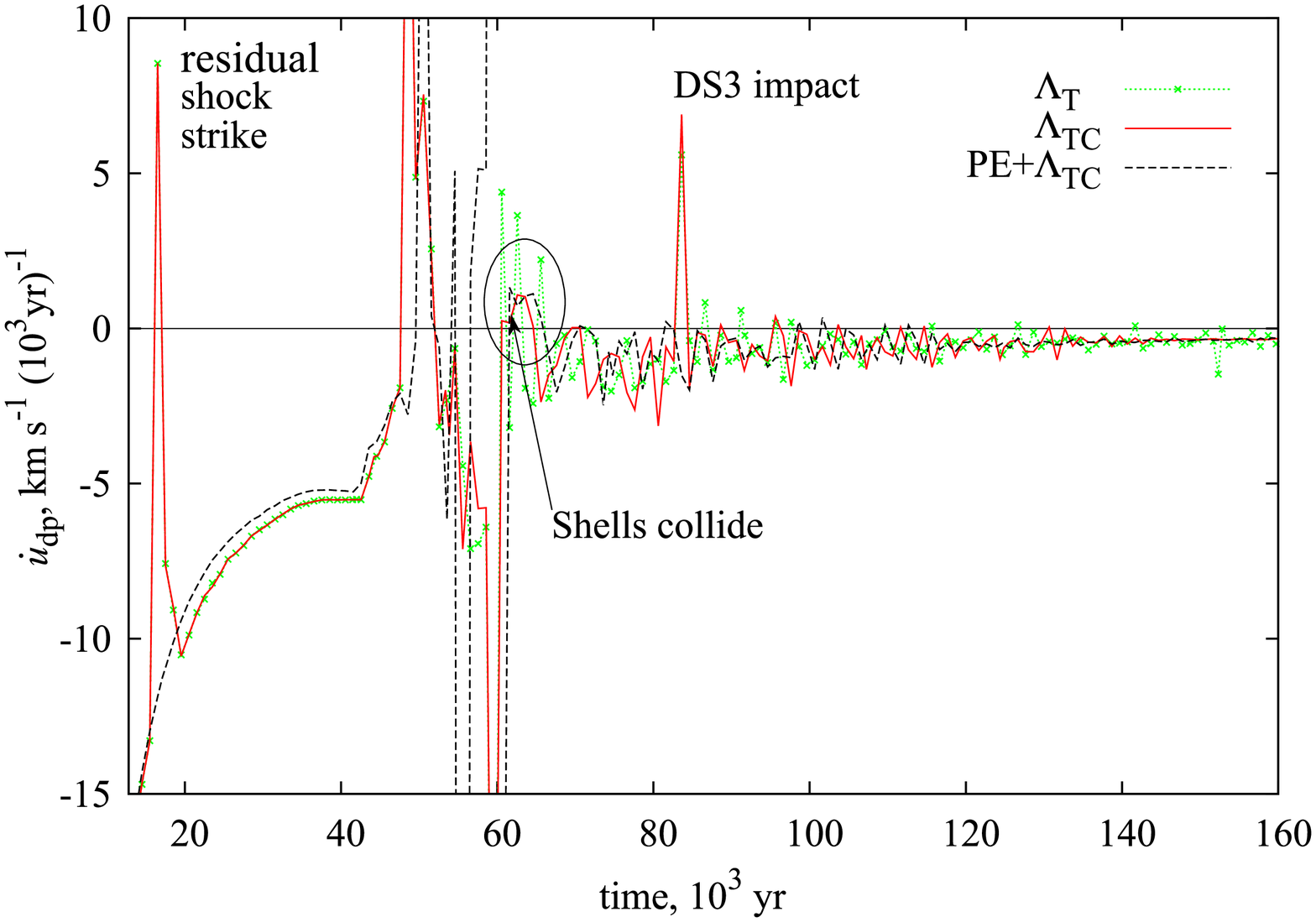}
  \caption{\label{fig:Accel}Time-evolution of the fluid velocity time-derivative at the highest density peak in different high-resolution E3-models: $\Lambda_\text{T}$ -- E3T-C19A, $\Lambda_\text{TC}$ -- E3TC-C19A, 
  $\text{PE}+\Lambda_\text{TC}$ -- E3TC-C19AP.}
\end{figure}

  We are going to illustrate the shell formation in detail with cylindrical explosions because their shocks decelerate slower and allow us to separate better key events 
 on plots, they are also relevant to our 2D calculations to be discussed further. See, e.g. Fig.~\ref{fig:StructC}, showing pressure and density profiles at various stages of the dense 
 shell formation. It is also useful to compare this with a set of `diagnostic' profiles of pressure, density, temperature, fluid velocity  as
 well as propagation speeds of the most interesting surfaces (forward shock jump `FS' and the highest density peak `dp') in Fig.~\ref{fig:Diagnostics}. Corresponding velocity derivatives are plotted in 
 Fig.~\ref{fig:Accel}. Strictly speaking, as the density peak position is a phase surface, these derivatives are not either partial (Eulerian) or substantial ones, 
 however since $60\times10^3$~yr the peak begins to move with the fluid, so $\dot{u}_\text{dp}$ can be considered as a Lagrangian acceleration of matter of the dense shell. 
 
 Curves in Fig.~\ref{fig:Diagnostics} correspond to three models with the highest effective resolution: E3T-C19A (full Table-CF), E3TC-C19A 
 (suppressed low-temperature cooling) and E3TC-C19AP (a point-like explosion, PE, its importance is reasoned below). They used mesh refinement and hence the apparent resolution varied in different
 parts of the flow, however, results of these models (like shocks and shells positions at different moments) are mostly consistent with those of fixed-mesh models E3T-C18 and E3TC-C18. 
 Nevertheless, several important distinctions will be pointed out below. 
 
 The state variables $\rho$, $p_\text{dp}$, $T$ and $u$ are taken directly from output tables (with $10^3$~yr time-step), while
 forward shock pressure $p_\text{FS}=\frac{2}{\gamma+1}\rho_0D_\text{FS}^2$, speed $D_\text{FS}$, the speed of the highest density peak propagation $\dot{R}_\text{dp}$ and 
 corresponding accelerations are derived by differencing corresponding data sequences (second-order centred differences). 
 
 It also should be kept in mind, that we allowed the ISM to cool in all these simulations, that is why its thermal
 pressure was always negligible. Otherwise, the hot ISM would introduce an additional scale of pressure or energy density, which would complicate generalizations between models with different explosion energies.

 Let us review the flow evolution from the end of the adiabatic stage until the reverse shock reflection and a little beyond. 
 
 Analysing behaviour of $t_\text{c}$ at the forward shock we have found that equating $t_\text{ad}=2t$ gives transition point estimates $t\approx 34\times10^3$~yr and $R\approx25$~pc. 
 Before that time the flow in general follows the adiabatic solution $R_\text{s}\sim t^(1/2)$.
 The forward shock is strong with the compression ratio 4, the density peak follows it, its material velocity is $\frac{3}{4}D_\text{s}$. 
 
 However, there were several moderate amplitude pressure and density perturbations observed behind the forward shock (see a smooth bump near 26.3~pc on the 44.4-curve in Fig.~\ref{fig:StructC}a). 
 These are residual pressure and entropy waves survived since the primordial discontinuity decay and multiply reflected from the centre and the forward shock. The 2~pc radius of the initial energy release domain 
 (see Table~\ref{tab:models}, this choice was imposed by much lower resolution of the 2D models to compare with) appears too large to allow these waves to damp before the catastrophic cooling begins. For example, one 
 of them catched up the FS at $17\times10^3$~years and provided a certain kick to it. Additionally, for a short time just before the collision the perturbed density exceeded that of the post-shock fluid, so that the 
 global density peak search pointed at the perturbation which resulted in a sharp `dp'-velocity change around $16-20\times10^3$~yr (see Fig.~\ref{fig:Accel}). 
 
 The existence of those reverberations means that the shock actually was not in complete dynamical balance with its interiors. They also suffer a certain dissipation while travelling between the shock and the centre,
 and it appear stronger in AMR-simulations than in regular-mesh ones, possibly due to inaccuracies of adaptation rules. 
 That is why $R(t)$-curves of E3T-C19A and E3TC-C19A display a lag with respect to point-like E3TC-C19AP  (see Fig.~\ref{fig:Diagnostics}c) while those of fixed-mesh E3T-C18 and E3TC-C18 almost coincide with it 
 (the lag is nearly ten times smaller, therefore the latter two are not shown).  
  
 It is more important, however, that such perturbations, though small, give rise to high density sub-shells later on when the intensive cooling begins (in all high resolution models with the 2~pc `thermal bomb' domain). 
 The point-like explosion simulation E3TC-C19AP has shown no `echo' and no perturbation induced fragmentation of the flow. 
 Except for the issues mentioned above, characteristic properties of the point-like explosion are in general consistent with other E3TC-models.
 
 A little after $4\times10^4$~yr when the FS runs beyond 29~pc (and corresponding $t_\text{c}$ decreases to $0.25 t$) 
 the divergence from the adiabatic law becomes evident and the catastrophic cooling begins (see the temperature curves in Fig.~\ref{fig:Diagnostics}). 
 
 Behind the forward shock the pressure gap starts to develop accumulating the cooling fluid around its bottom.  
 The global density peak also departs from the shock and transfers there, see a divergence of $R(t)$ curves in Fig.~\ref{fig:Diagnostics}c and an abrupt fall of 
 $\dot{R}_\text{dp}$ in Fig.~\ref{fig:Diagnostics}d at $42-43\times10^3$~yr. 
 
 While the peak position moves deeper with respect to the FS (but still outwards in the lab frame as $\dot{R}_\text{dp}>0$), 
 the fluid velocity $u_\text{dp}$ at it does not display sharp features and even exceeds that of the FS jump ($\frac{3}{4}D_\text{FS}$). 
 During this time the shell compression does not exceed significantly that of the forward shock (see Fig.~\ref{fig:Diagnostics}a). 
 
 However, around $49-50\times10^3$~yr the pressure gap becomes an order of magnitude deep (one can compare the FS pressure to that of the density peak in Fig.~\ref{fig:Diagnostics}b, 
 see also Fig.~\ref{fig:StructC}a), and its strong left gradient starts to re-accelerate the cooling fluid outwards: 
 see the bumps of $u_\text{dp}$ and $\dot{R}_\text{dp}$ in Fig.~\ref{fig:Diagnostics}d and the sharp spike of positive acceleration in Fig.~\ref{fig:Accel}. 
 
 This `jump' coasts near $52-53\times10^3$ in all models as thermal pressure in the peak exceeds that of the surroundings (see the local pressure maximum in Fig.~\ref{fig:StructC}a near 28.8~pc). 
 This is achieved both by rapid compression (five times per thousand years) and the cooling slowing down as the temperature falls considerably below the $\Lambda_\text{T}$-maximum (i.e. to several thousand K, 
 see Fig.~\ref{fig:CF}). Additionally, while the dense shell accelerates, the forward shock -- in contrast -- loses its speed, therefore the forward cooling region to the right of the shell 
 contracts quicker than the rear one (to the left). 
 
 The density peak quickly departs from the forward shock (compare $R_\text{dp}$ and $\dot{R}_\text{dp}$ with $R_\text{FS}$ and $D_\text{FS}$ 
 in Figures~\ref{fig:Diagnostics}c-d within $42-50\times10^3$~yr interval) and after that moves with the fluid velocity (compare $u_\text{dp}$ and $\dot{R}_\text{dp}$ after $50\times10^3$~yr). 
 Meanwhile the shock speed decreases, so does the post-shock temperature (and the corresponding cooling time), and eventually the it appears separated from the density peak by more than one 
 cooling length. At this time a second dense shell (hereafter DS2, and the first one -- DS1) should form, which is clearly seen in Fig.~\ref{fig:StructC}a near 29.6~pc. 
 We call it `spontaneous shell production', in contrast to the induced by the `echo'-perturbations.  
 
 Such self-repeating thermal instability has been already described in \citet{Falle1981}, however in that paper it was interpreted as a result of collision of an inner shock 
 (a bounce-shock produced when the shell contraction terminates, see figure~5 of that paper) and the forward one. 
 Oppositely, in our simulations the secondary shell is triggered not by any inner shock but by a rarefaction wave (it is more proper to call it an unload wave)
 reaching the forward shock (around $t\approx53-55\times 10^3$ years), lowering its speed, temperature and cooling time. The rarefaction appears due to matter exhaustion towards the primary shell well 
 before than the it contracts to its minimal thickness. A similar explanation ($t_\text{c}$ shortening) for double shell production was also put forward in \citet{Blondin98}. 
 
 In the `echo'-free E3TC-C19AP model the second peak even overgrows the first one for a certain time, so the peak-tracking routine switches to it from $55\times10^3$~yr to $59\times10^3$~yr 
 (resulting in sharp jump up and down in $\dot{R}_\text{dp}$ seen in Fig.~\ref{fig:Diagnostics}d, in Fig.~\ref{fig:Diagnostics}c it is labelled with `PE: dp at DS2') 
 and follows the forward shock closely (as the secondary peak is situated just behind it). 
 
 It should be mentioned that this second shell also experiences an episode of acceleration which drives forward shock revival. The growing FS-pressure then 
 coasts the second shell material velocity and even stops for a while its peak propagation. Therefore a sharp drop of $\dot{R}_\text{dp}$ below 50~km~s$^{-1}$ at $58\times10^3$~yr (Fig.~\ref{fig:Diagnostics}d) is of physical and not of numerical or 
 algorithmical origin: the DS2 density profile becomes more asymmetric while it accumulates matter from its rear at a higher rate than from the front, its point of maximum moves away from the FS although the material velocity at that point is close to 
 the forward shock speed. 
 
 The first shell approaches the second one which is slower and they exchange momentum via the overcompressed forward cooling region between them like two bodies linked by a spring (although with decreasing stiffness). 
 The former has greater inertia (it is clear from the $\rho u_\text{dp}^2$ of the point-like explosion curve in Fig.~\ref{fig:Diagnostics}b when it switches to DS2 at $55\times10^3$~yr, the accumulated mass of DS1 is also higher) and
 its relative velocity loss is less than the relative gain to the latter.
 
 Obviously, the effective cooling length shortens not only at the forward shock, but -- even quicker -- in the first dense shell too. 
 So the latter fragments into a growing number of sub-shells (which is seen at the left slope of DS1 in Fig.~\ref{fig:StructC}a).
 This is enhanced by non-monotonicity of $\Lambda_\text{T}(T)$ near its maximum: there are several regions of the shell that cool faster, contract faster, the cooling length decreases more and more and the process
 repeats on finer scales.  They stay separated for a while due to dispersion of velocity gained by different elements of the fluid being accelerated by the pressure gap slope. Then they subsequently merge one with another and with the main body of the DS1. 
 
 After a detailed comparison of different models, we have concluded, however, that this multiscale fragmentation is caused purely by numerical noise of interpolation procedures of mesh refinement. The flow at that time appears unstable and susceptible to
 unphysical variations of pressure and density. Nothing like this was observed in both regular-mesh simulations E3T-C18 and E3TC-C18, so one should be careful about this kind of artefact-induced fragmentation when performing high-resolution AMR calculations
 (especially multidimensional ones). 
 
 As the numerical noise can produce such a pronounced effect, much more vigorous deviations should be expected from artefacts like residual waves (or the initial explosion reverberations).
 
 Indeed, by occasion, in models with 2~pc thermal bomb domain (all -C18 and -C19A ones with exception of point-like E3TC-C19AP) there was one of them situated near the first shell birth place. 
 This appeared to enhance the compression significantly and resulted in a quickly growing sharp spike sticking out from a wide bump (as it is seen in Fig.~\ref{fig:StructC}a around 29-30~pc). 
 
 The spike lay a bit behind the bump maximum, so once it had risen above the latter the peak-seeking routine became to trigger on it immediately, and that is why a short step backwards in $R_\text{dp}$ and a sharp drop of 
 $\dot{R}_\text{dp}$ show up in Figures~\ref{fig:Diagnostics}c-d at $49\times10^3$~yr (E3T-C19A and E3TC-C19A models, $R_\text{dp}(t)$ of E3TC-C19AP exhibits a similarly-looking step, but it was caused by a small noise induced sub-shell). 
 
 After that moment this spike was always the highest one and therefore all `dp'-values were taken only from it (in all models excepting the point-like E3TC-C19AP again). 
 This explains the divergence seen in Fig.~\ref{fig:Diagnostics}d between `dp'-velocity curves of the point-like explosion and two other models within $48-60\times10^3$~yr time interval. 
 
 By $60\times10^3$~yr DS1 again becomes the densest even in E3TC-C19AP (see the label `PE: dp at DS1' in Fig.~\ref{fig:Diagnostics}c), and finally merge with DS2 at $61\times10^3$~yr ($60\times10^3$ in the two other models).
 
 In Fig.~\ref{fig:StructC}a the dashed black arrows show schematically trajectories of the first and the second shells, from their birth until coalescence into a single peak. 
 There are two spontaneously produced shells and at least two resulting from the `echo'-waves. 
 
 Although in that figure only profiles of the E3TC-C19A model are shown, one can also estimate how look the others. All models agree well in reproducing times and radii of key episodes of dynamics of the dense shells. 
 All non-point-like explosions give two reverberation-induced spikes (the second of them is labelled as DS3 formed in front of the expanding interiors), 
 while E3TC-C19AP produces no other spikes beside DS1 (its trajectory is depicted in Fig.~\ref{fig:StructC}a by corresponding arrowed line) and DS2.
 
 All AMR-simulations generate the noise-induced substructure, while the regular-mesh ones do not and their DS1 and DS2 have smooth monotoneous wings. Additionally, reverse shocks in both E3T-C18 and E3TC-C18 are also more regular and flat than in AMR, 
 pulsations of the forward shock speed are weaker (130-180 and 100-190~km~s$^{-1}$ for E3T-C18 and E3TC-C18 respectively) and damp faster (by $90\times10^3$~yr).
 
 Models with the unrestricted table cooling function yields higher shell densities, lower temperatures, later times of the reverse shock separation ($90-95\times10^3$~yr instead of $70-75\times10^3$~yr), 
 their reverse shock profiles are also smooth and flat, being indicative of a greater role of numerical viscosity.
 
 \subsubsection{Structure of the dense shell and pre-conditions for multidimensional instabilities}\label{sec:1Dstruct}

  \begin{figure}
  \centering
  \includegraphics[width=1.\linewidth]{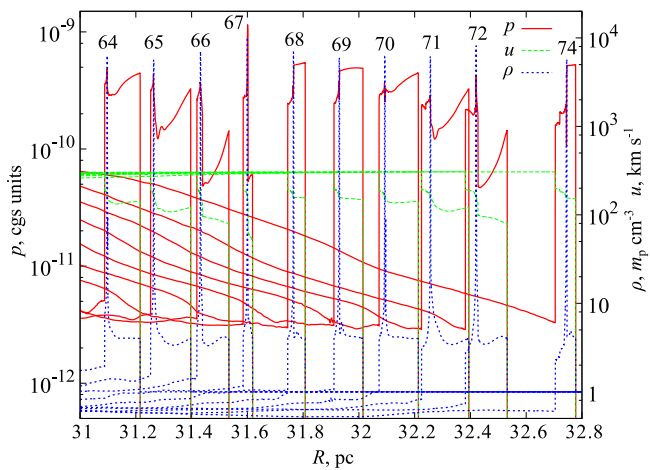}
  \caption{\label{fig:Puls} Pressure ($p$), fluid velocity ($u$) and density ($\rho$) profiles illustrating the flow structure between the two shocks during two periods of pulsations in the E3TC-C19A simulation.}
\end{figure} 

 \begin{figure*}
  \centering
  \includegraphics[width=1.\linewidth]{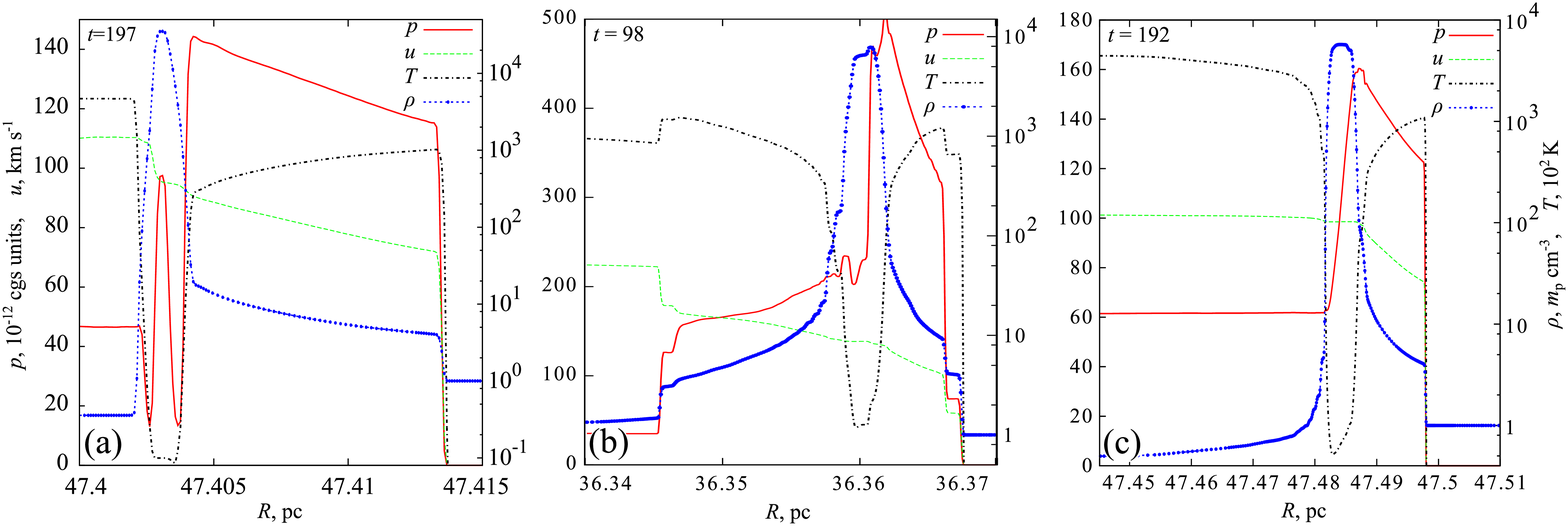}
  \caption{\label{fig:Fine}Profiles of pressure $p$, fluid velocity $u$, temperature $T$ and density $\rho$ in the closest vicinity of the dense shell calculated with different cooling functions. 
  (a) -- E3T-C19A with full unconstrained Table-CF, 197 thousand years. (b) -- E3TC-C19A with the constrained cooling function $\Lambda_\text{TC}$, 98 thousand years. 
  (c) -- the same for 192 thousand years. }
\end{figure*} 
 
 During the shell merger the peak density reaches its extremum (see e.g. corresponding profiles in Fig.~\ref{fig:StructC}a or evolution curves in Fig.~\ref{fig:Diagnostics}a).
 The resulting shell (hereafter simply the shell) accumulates almost all of the shocked fluid mass and momentum. It is catching the `stagnated' forward shock up quickly. 
 So, it is far from balance with its environment, and there immediately begins a (long) process of dynamical equilibrium establishing between the shell, the forward shock and the hot interiors.
 
 At this stage of primary importance is how pressure and momentum are transported by internal waves and what stops further contraction of the shell.
 
 Just after the collapse, the shell still continues to gain velocity (it is seen in Figures~\ref{fig:Diagnostics}d and \ref{fig:Accel}) absorbing a trailing `tail' of the cooling fluid 
 previously accelerated at the pressure gradient. Then the acceleration is terminated by growing pressure between it and the forward shock. 
 
 The shell quickly compresses the forward cooling region (driving even an additional internal shock through it) and this `kicks' the forward shock. While the shell 
 feels a certain recoil, the speeded-up forward shock re-heats matter which increases the loss rate, and again a new pressure drop develops behind it (this somewhat releases a load upon cold 
 fluid) so the forward cooling region falls back on to the dense gas. One more internal shock is reflected into the FCR, and the shell suffers one more episode of momentum tapping.
 
 To its rear, a similar sequence of events takes place, however with that difference that matter there is hotter, less dense and moves faster than the shell (so the reverse shock can be also named
 the termination shock).  
 
 In series of such pulsations the rear and forward cooling regions push and pull the shell one to another until the three appear matched in ram and thermal pressure. 
 Excessive kinetic energy of the shell is dissipated via enhanced entropy production at shocks and consequent radiation. This is clearly seen in Figures~\ref{fig:Diagnostics}c-d. 
  It is often referred to as a radial overstability and is reliably observed in different 1D simulations \citep[see e.g.][]{Falle1981,Blondin98}.
 
 Fig.~\ref{fig:Puls} illustrates two periods of such pulsations. Profiles at 68 and 74~$10^3$~yr 
 correspond to the phase of compressed and reheated FCR, after which the FS departure is seen, as well as consequent cooling and falling back.  
 One can see that during the latter stage even a weak sub-shell may form within the contracting cooling region, however is was not observed
 in regular-mesh simulations and therefore it is likely to be triggered by mesh-refinement perturbations.
 
 Less obvious is how the modelled shell sustains its shape, withstands high pressure and transmits waves from one cooling region to another. 
 In ideal hydrodynamics there are only two possibilities: the thermal pressure $p_\text{dp}=nkT$ and numerical or scheme viscosity. 
 In real flows a non-thermal supporting force may be provided also by magnetic fields \citep[and may appear dominant even before the final shell formation, see][]{PetrukKuzyoBeshley}. 
 
 As it is seen from Fig.~\ref{fig:Puls}, during the pulsations the forward cooling region pressure is of order 
 \begin{equation}
   p_\text{FS}=\frac{2\rho_0}{(\gamma+1)}D_\text{FS}^2=1.25\times10^{-10}D_\text{[100]}^2\text{~}\text{ cgs units},\label{eq:pshock}
 \end{equation}
 where $D_\text{[100]}$ is the forward shock speed in km~s$^{-1}$. On the other hand, according to the adopted equation of state the thermal pressure at the density peak is
 \begin{equation}
  p_\text{th}=\frac{\rho}{\mu m_\text{p}}kT=2.76\times10^{-10}\rho_\text{dp,[4]}T_\text{dp,[2]}\label{eq:ptherm},
 \end{equation}
 where $\rho_\text{dp,[4]}$ is the peak density in $10^4$~$m_\text{p}$~cm$^{-3}$ and $T_\text{dp,[1]}$ -- its temperature in 100~K. 
 
 Substituting corresponding values from Fig.~\ref{fig:Diagnostics},
 one can easily see that with the suppressed table cooling function $\Lambda_\text{TC}$ there is still enough heat in the simulated shell to sustain the forward shock by means of thermal pressure. 
 We have also found that even at the very peak (i.e. the location of the highest loss rate) a size of `hydrodynamically causally connected' domain 
 $\lambda_{hc}=(t_\text{c}c_\text{s})|_\text{dp}$ after the shell coalescence grows from $\sim10^{-3}$~pc up to $\sim10^{-1}$~pc. During the same time the resulting shell
 width (at the base) increased from 0.003~pc up to 0.01~pc while its density somewhat decreased. 
 
 Therefore we believe that the shell is well resolved on our C19A meshes with the finest available cell width of $9.5\times10^{-5}$~pc and the simulations with the Cut Table-CF 
 $\Lambda_\text{TC}$ reproduce more or less physically correctly (as much as $\Lambda_\text{TC}$ is a good approximation) how the shell shape is supported and how the waves are transmitted through it.
  
 Oppositely, models with the unrestricted cooling $\Lambda_\text{T}$ never converged to a self-sustained shell (we tested this with resolution up to $2^{18}$ regular-mesh cells  and 
 $2^{19}$ finest AMR cells per 50~pc), its temperatures held near the minimal allowed limit (10~K), $\lambda_\text{hc}\sim10^{-6}$~pc (in the highest resolution run E3T-C19A) and the thermal pressure
 always displayed a prominent drop at the density peak. This means that the shell was supported by mostly by numerical or scheme viscosity.
 
 Unlike an artificial viscosity usually introduced
 explicitly as an additional term in equations \citep[this was used in Lagrangian simulations of][and that is why an infinite compression 
 was not obtained there even despite unconstrained cooling]{Straka}, the scheme viscosity is just an inherent property of discretized Euler equations due to which every jump, 
 and a density spike in particular, should be inevitably smeared over not less than a certain number of cells depending on a chosen numerical scheme.
 
 This limits the simulated dense shell compression ratio (in a homogeneous medium) by
 \begin{equation}
   \eta_\text{lim}=\frac{R}{N_\text{D}H_\text{lim}}=\frac{R}{R_\text{max}}\frac{N_\text{tot}}{N_\text{D}N_\text{lim}}\label{eq:etalim},
 \end{equation}
 where $R$ is a current position of the dense shell (which is assumed to accumulate all the swept up mass), $N_\text{D}$ -- number of symmetry dimensions (2 for cylindrical symmetry and 3 for spherical one),
 $R_\text{max}$ -- full size of computational domain,
 $N_\text{tot}$ -- total number of cells in it (or equivalent number of the finest cells in cases of AMR), $H_\text{lim}$ and $N_\text{lim}$ -- minimal equivalent width of a spike (in length units and
 in pixels, respectively) allowed by a chosen scheme with given resolution and number of cells (the finest available) within it. 
 
 $N_\text{lim}$ assesses ability of a reconstruction rule, used in the scheme, to reproduce an essentially
 non-linear behaviour of the spike profile (i.e. requiring high order expansion terms). 
 From our experiments we can roughly estimate that the monotonicity preserving second-order accurate MUSCL-scheme (used in \small{FRONT3D})
 requires $N_\text{lim}=5-6$ cells near the peak top, while the third-order PPM (used in \small{FLASH}) -- $N_\text{lim}=3-4$. 
 
 Having $\eta_\text{lim}$ estimated, one can judge, comparing $\eta_\text{lim}$ to the calculated compression, whether the scheme viscosity is important in obtained results. On the other hand,
 if one expects the forward shock to have some certain speed (and therefore the post-shock pressure) and also the characteristic temperature of the cooling cut-off is known, this criterion can help
 to chose a proper grid spacing in order to avoid significant scheme viscosity: 
 \begin{equation}
   \delta R < \frac{R_\text{FS}}{C_\text{sc}N_\text{D}N_\text{lim}}\frac{(\gamma+1)kT_\text{cut}}{2\mu m_\text{p} D^2_\text{FS}}\label{eq:limdr}
 \end{equation}
 where $C_\text{sc}$ is a scaling constant of order of at least several. 
 
 In our simulations we had $\eta_\text{lim}\approx1.3\times10^4$ for the C18-models and $\eta_\text{lim}\approx4.5\times10^4$ for the C19A ones. From Fig.~\ref{fig:Diagnostics}a one can clearly see,
 that unconstrained cooling simulations clearly do not meet the criterion \ref{eq:limdr}, while those with $\Lambda_\text{TC}$ do. 
 
 These considerations can be well illustrated with Fig.~\ref{fig:Fine} showing structure of the shell closest vicinity at late stages of its evolution. 
 The lack of thermal pressure within the unconstrainedly cooling shell is evident (panel (a))\footnote{Also, the lack of balance between the interiors and the forward shock results
 in that the latter is driven only by the shell inertia and its radius converges to the moment conserving solution rather than PDS. This should be common for all un-resolved shells}, 
 while from two other panels it is seen that the shell comes to dynamical equilibrium with the 
 forward and rear cooling regions by exchanging waves (see pressure variations and internal shocks in Fig.~\ref{fig:Fine}b) and then relaxes to a quasi-stationary 
 self-similar structure (Fig.~\ref{fig:Fine}c).

 This structure differs from an analytical solution obtained for two colliding equal supersonic flows with constrained radiative cooling which was obtained by \citet{Creasey} 
 (see figure~1 of that paper). Definitely, this is because of that our shell moving with rather high velocity, which means that there is an asymmetry in ram pressure between the forward and 
 rear cooling regions. 
 
 This obviously can not be reduced by transformation into the shell comoving frame, and makes our case somewhat differ from the setup used to derive the 
 NTSI~\citep[see][]{Vishniac1994}. So, from Vishniac's non-linear instabilities, it is NDI that seems relevant for RCSs, although its analytical theory is not as developed as that of NTSI.
 Additionally, in non-resolved simulations, when the shell is overcooled, the bi-directional mass and momentum transfer, which is the `engine' of NTSI, must be somewhat suppressed by the 
 reduced wave-transmittancy of the shell due to low speed of sound so that perturbations should stick within it rather then deform the opposite site. The transfer, however, still may be possible
 via the scheme viscosity. 
  
 Beside of being potentially Vishniac-unstable just due to the two-shock confinement (during pulsations there are also internal shocks which may enclose it), the shell is also Rayleigh--Taylor-unstable.
 The first evidence to this is appearance of the transient 
 acceleration episodes (see~Fig.~\ref{fig:Accel}) during which the shell feels certain `kicks' from below (the impulsive acceleration makes this also relative to the Richtmyer--Meshkov instability). 
 The second -- the fact that the pressure gradient at its forward face is directed oppositely 
 to the density gradient (see Fig.~\ref{fig:Fine}) making this region possibly Rayleigh--Taylor unstable as well. 
 
\subsection{The 2D shell evolution and instabilities}\label{sec:2Dinst}

 Taking into account a number of instabilities which have to take place in the RCS thin shell (the thermal instability, noise-induced fragmentation, Vishniac and Rayleigh--Taylor instabilities), it is not
 surprising that the shell appears unstable against even numerical perturbations caused by misalignment between lines of flow and grid.

 In all three used codes, when simulated on Cartesian 2D grids with the Table-CF, the RCS thin shell after running nearly 0.3-0.5 of its initial radius smoothly and symmetrically then begins to bend and 
 subsequently breaks up. The circular shaped shells are radial outflows and on polar meshes they are reproduced correctly (in quantitative agreement with 1D-results). 
 This gives us a possibility to go further and investigate their
 response to physical perturbations. In following subsections we are going to discuss both these cases.

 \subsubsection{Bending instability on Cartesian grids}
 
\begin{figure*}
 \includegraphics[width=0.95\linewidth]{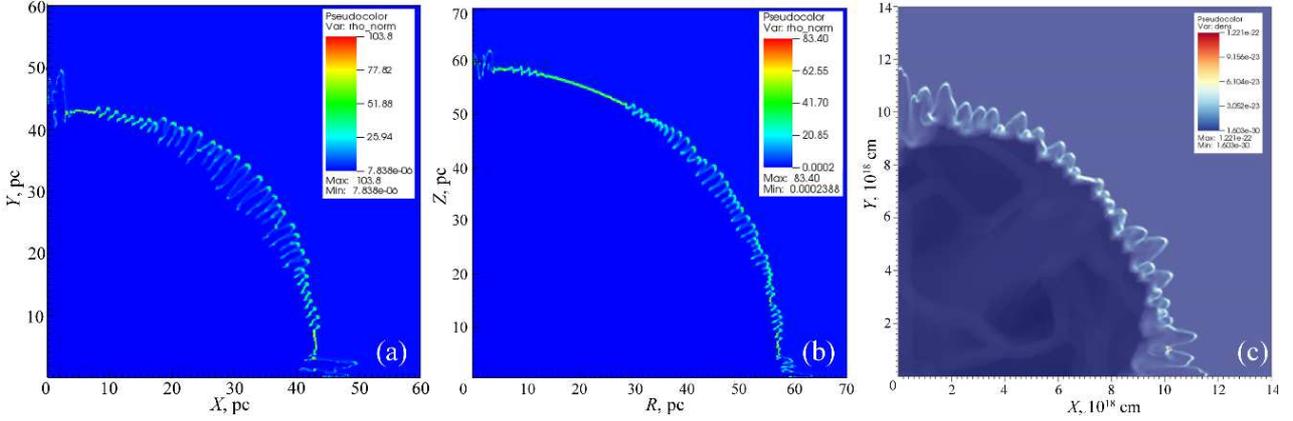}
  \caption{\label{fig:XYCurls} A typical look of the developed thin shell bending instability in cylindrical simulations on Cartesian grids. Density profiles: (a) -- cylindric E3T-XY9 and (b) -- axisymmetric E1T-RZ9 in $m_\text{p}$~cm$^{-3}$,
  (c) -- E4T-XY9A in g~cm$^{-3}$.}
\end{figure*}

In order not to overload our narration with numerous similar-looking pictures we have placed the most of 
figures, illustrating the instability development in detail,
into Appendix~\ref{app:InstEvol}, additional high-resolution large-scale pictures and videos are available online and through our group's web site \url{http://dau.itep.ru/sn/radshocksnr/}. 

Fig.~\ref{fig:XYCurls} shows the bending instability as it typically appears in our calculations. Similar looking patterns show up in works of other authors \citep[e.g.][]{KimOstriker14}.

In planar symmetry (XY-setups) the `curls' are nearly symmetric with respect to a bisecting line (inclined by $45^\circ$ to the grid axes) on regular fixed meshes ({\small FRONT3D} and
{\small PLUTO4}), but on the refined mesh of E4T-XY9A the symmetry is not so perfect. In axisymmetrical $RZ$-calculations the bends look identically (Fig.~\ref{fig:XYCurls}b), 
but are somewhat suppressed at low polar angles: especially whithin the range of $\theta=10-35^\circ$, at other latitudes their amplitudes are similar to those of $XY$-simulations. 

On fixed meshes, soon after the thin shell forms, the first density fluctuations and 
ripples appear near the domain boundaries. This is likely to be due to the carbuncle numerical instability (they are more pronounced in pure HLLC-calculations of \small{FRONT3D} as we did not introduce
any special artificial dissipative correction to them). They are quite localised and do not propagate far along the shell limb. 

After that, around the bisector there come advancing bends overrunning the main shock. Then appear series of additional arcs: 
the further from the bisector -- the later. Their initial growth rate is high, with an excess velocity as high as $(0.3-0.5)D_\text{s}$. However, it quickly saturates, thus somewhat 
later the new arcs generation cancels, and their pattern evolves rather slowly, mostly scaling with the average shock radius. Matter of the shell concentrates 
predominantly in the leading arcs, though there are also several trailing ones with high-density clumps. 

\begin{figure*}
   \includegraphics[width=0.95\linewidth]{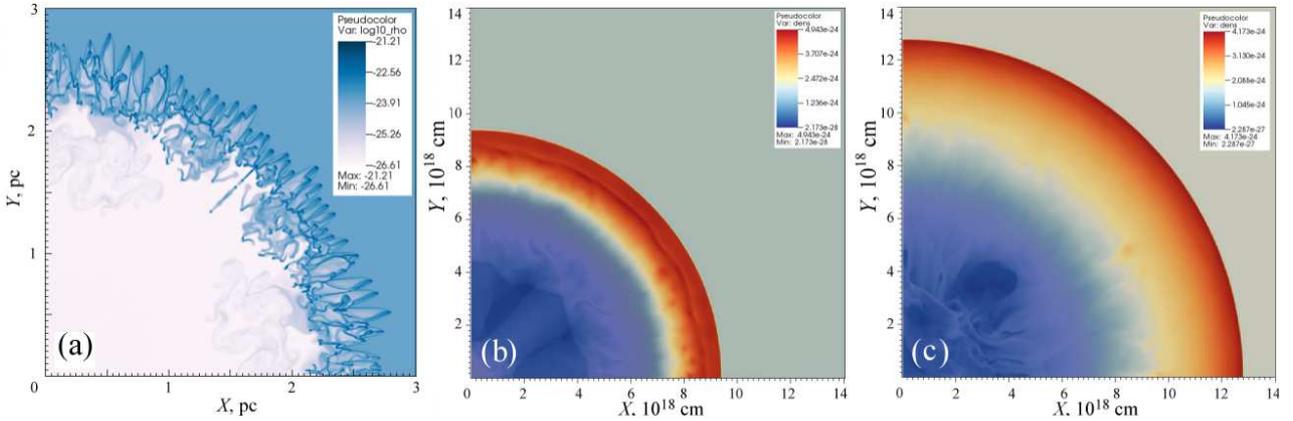}
  \caption{\label{fig:LateStraka} (a) -- logarithmic density profiles (in g~cm$^{-3}$) of the latest stages of the E4T-XY10 cylindrical explosion. (b),(c) -- intermediate and late stages of the dense
  shell smearing with Straka-CF cooling in the E4S-XY9Asimulation.} 
\end{figure*}

In compact low energy models E4T-XY9A, E5T-XY9A on adaptive meshes with moderate refinement levels (see Fig.~\ref{fig:E4TAMR},~\ref{fig:E5TAMR} in Appendix~\ref{app:InstEvol}), contrary, 
the rising arcs appear almost simultaneously with `carbuncles' and far from the bisector. 
It is seen that the shell appears somewhat smeared out, and `drops' or clumps of increased density form within it, then they move ahead and produce the arcs. 

In cases of increased ambient density (which means a slowed shock and two orders of magnitude enhanced cooling) and lowered $\gamma$ 
(increased compressibility) even more complex flow dynamics is observed (Fig.~\ref{fig:E5TAMR} in Appendix~\ref{app:InstEvol}).
The drops are likely to form from mushroom-looking structures behind the RCS front 
collapsing due to the catastrophic cooling. Once it happens, the heavy clumps, decelerating slower than the rest of the shocked fluid, proceed to the shell's forward 
surface and then strike and penetrate it. During this time the RCS radius approximately doubles. One also may notice that in this case matter stays mostly within the shell and
not in the advancing arcs. 

At the latest stages, the whole RCS flow (except for the inner hot gas) separates into a set of `micro-jets' consisting
of dense cores driving their own individual bow shocks (see Fig.~\ref{fig:LateStraka}a). Inside the hot bubble, there are bent loops, plumes and `mushroom'-looking patterns of density
clearly seen, resembling those of the RT-instability.

\subsubsection{The special cooling treatment does not matter}

\citet{Townsend_ApJS_2009} has shown, 
that general-purpose non-linear ordinary differential equation solvers, either explicit or implicit
(e.g. based on the Newton--Raphson method) ones, widely used for cooling rate integration, 
sometimes might be ambiguous, unstable and converge to very different solutions from similar initial conditions.
This may occur when
the cooling function has a complex and essentially non-linear multipeaked temperature dependence (see e.g. our 
Table-CF in Fig.~\ref{fig:CF}).
This would potentially result in spurious temperature and pressure gradients, which would drive unphysical flows. To avoid these drawbacks, a special exact algorithm to account  for
the non-linear cooling has been described by~\citet{Townsend_ApJS_2009}.

However, the observed instability appears independent on whether this
exact algorithm is used (like in the {\small FRONT3D} code,
see also Appendix~\ref{app:CoolTown}) or not (all three codes can utilize the Newton--Raphson solver), 
and this does not explain the instability of RCS with the complex Table-CF. Moreover, in some cases it was observed even with rather regular Straka's and monotonic power-law cooling functions 
(e.g. in E2S- and E6F-explosions), however it took considerably larger time for the bending to develop.  

Therefore, we conclude that for the RCS thin shell global modelling one can still use general-purpose codes for it.

\subsubsection{The instability growth dependence on cooling function}

The instability is more easily excited with the complex multipeaked Table-CF than with the Straka-CF or free--free power law. One can easily guess that this is because the Table-CF implies shorter 
cooling time-scales, higher pressure gradients and accelerations.

It was found, however, that under certain initial conditions (lower $E_0$ and/or higher $\rho_0$) and less violent cooling functions (Straka-CF), the perturbations can decay 
and the dense shell -- smear out not very long after it had been formed. Likely, the temperature had quickly settled down to the power law part of Straka-CF and the cooling became slow 
($t_\text{c}>t_\text{ad}$) resembling an adiabatic flow but with a lower effective $\gamma$.

We illustrate this with Fig.~\ref{fig:LateStraka}b-c for the E4S-explosion with AMR. There are some clumps forming behind the shock, and even a certain secondary shell is seen, 
but the former ones then smear out and the latter -- catches the forward shock up and disappears. A similar picture was observed in the fixed-mesh E4S-XY8 run. 
This can be easily understood if one recalls that 
the E4 and E5 models are the short-scale ones (5~pc for the whole domain), which means shorter dynamical time-scales. Obviously, this facilitates the catastrophic cooling stabilization, as $t_\text{c}$
does not deviate from $t_\text{as}$ so considerably like in the large-scale explosions. 

During the stabilization, in E4S-XY9Aruns the peak compression gradually decayed from its maximum (nearly 25) to $\approx4$, 
suggesting that the flow relaxed to an adiabatic regime with $\gamma\approx5/3$ like in initial conditions. 

In the `high-density' runs E5S (the same as the E5T-models from Table~\ref{tab:models}, but with the Straka-
CF) there were no sub-structures observed, the shell compression evolved quickly from 12 to 8.15 and then fluctuated with a span of 0.1 around the latter value 
until the end of the run. This is greater than the compression 6 at the strong adiabatic shock for $\gamma_{\text{E5}}=7/5$. 
This effective gamma lowering may be naturally explained by the 10 times higher density 
(with respect to E4S) and the 100 times higher cooling rate, though the latter is not catastrophic.  

It is also found that the low-temperature constrained cooling function $\Lambda_\text{CT}$ does prevent the bending, which means that the latter is determined just by the catastrophic cooling existence 
and not by properties of what it finally produces (e.g. the shell width, temperature or density). It is clear that resolution of 2D-calculations is always not sufficient to meet the criterion~\ref{eq:limdr} 
and for the constrained cooling to have an observable effect on the instability (unless the cut-off temperature is too high, of order of $10^4$~K).

\subsubsection{The bending on Cartesian grids is caused by the RT-like instabilities excited by numerical noise}

Here we put forward several arguments in favour that the observed effect is not any of the Vishniac instabilities but relative to Rayleigh--Taylor's one. 

The first argument is lack of resolution in 2D-models. The NTSI or NDI require the shell bounded by two shocks, but in large-span models (like E1-E3) the corresponding structure 
(forward shock -- dense shell -- reverse shock) is not resolved (actually a resolution comparable to that defined by equation~\ref{eq:limdr} is required, and this is rather demanding computationally),
and profiles look similar to those shown in Fig.~\ref{fig:StructS}a.  In the short-span E4T-XY12A simulation the rear cooling region is resolved well, the forward one somewhat worse, but is still 
distinguishable from the shell which is still unresolved. 

In addition, whatever the resolution, numerical noise perturbations (the only available in our simulations on Cartesian grids) always have wavelengths comparable to the cell width. 
Therefore the numerical viscosity, which is strong at such scales, obviously overdamps the Vishniac instabilities, and that is why the shell can travel nearly 30-50 per cent of its initial formation radius 
unperturbed (with except of the `carbuncles') before the first bends become pronounced due to another reason.

The second argument regards the shape of the bends. They appear rather suddenly, grow fast and then `freeze' their pattern. 
PDTSO-like oscillations were not observed. The perturbations do not resemble a typical NTSI pattern \citep[i.e. produced by collision of two winds: symmetric bends 
with density clumps at extrema due to mass accumulation and curls along slopes due to shear motions, gradually taking a saw-like shape and the shell becoming torn in opposite directions, 
see e.g.][]{BlondinMarks96, McLeodWhitworth13} as well, especially in AMR-simulations. Thus, unlike the NTSI predictions, there is mass transfer through the shell observed predominantly 
in only one direction, namely outwards.

Even in widely referred to high-quality simulations of the SNR adiabatic-to-radiative stage transition carried out by \citet{Blondin98}, 
although the obtained results are explained only with NTSI and PDTSO, in fact the late-time density 
patterns  there deviate from those of the `canonical' NTSI, there also is no clear symmetry between front and rear of the shock-bounded shell, but pictures display some typical features of the RT-like 
instabilities: rising bubbles and sinking rolled-up spikes. 

Third, in simulations with unresolved shell structure (namely, all two-dimensional E1-E3 ones and also E4-E5 with equivalent cell number less that 1024), the moment of the instability correlate with episodes 
of positive acceleration (see Fig.~\ref{fig:Accel}). However, the latter ones depend on resolution too: while in E3TC-XY11 there is indeed an episode of positive acceleration at $62-68\times10^3$~yr, like in 
Fig.~\ref{fig:Accel}, and the bending bursts out exactly at that time, but in E3T-XY9 (the resolution of which 512 cells per 80~pc seems now rather poor) due to high numerical viscosity the episode of 
acceleration and start of instability delay until 100 thousand years when the pressure gap behind the shell begins to close. The former episode is connected with the shell collapse and absorption of trailing 
cooling fluid in its vicinity (and this is not reproduced with low resolution), but the latter one is relative to the acceleration spike at 90 thousand years 
(labelled as `DS3 impact' in Fig.~\ref{fig:Accel}) which is clearly caused by impact of matter dragged by the pressure drop between the shell and interiors. 

\begin{figure*}
   \includegraphics[width=0.95\linewidth]{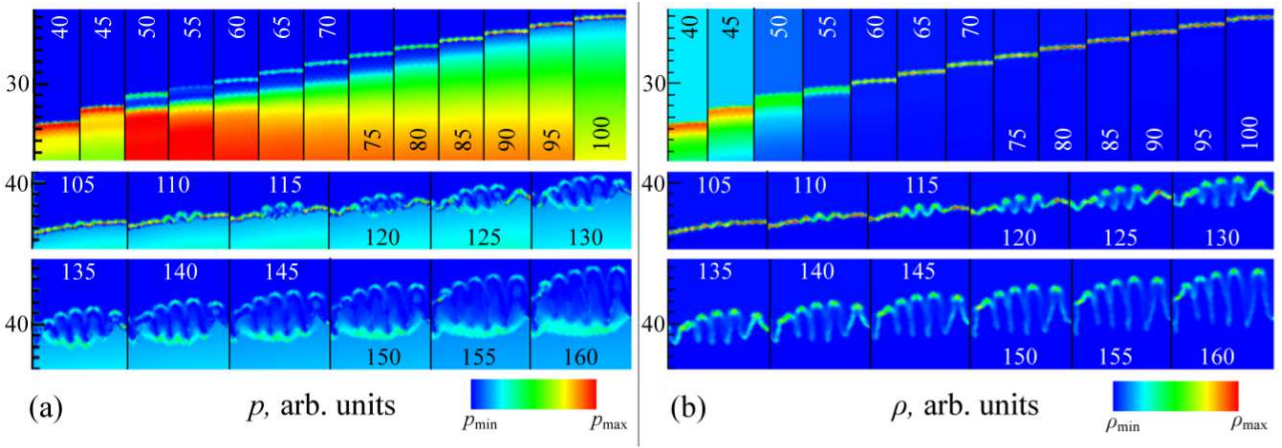}
  
  \caption{\label{fig:PRhoSeriesFront} Series of cuts of pressure (a) and density (b) fields of the E3T-XY9 run illustrating the shell formation and pressure driven instability development. Right boundaries 
  of each cut are aligned to the domain bisector (i.e. $45^\circ$ inclined to both axes), scales to the left are radii from the centre in parsecs, numbers within the cuts indicate time in $10^3$~yr. 
  Pressures and densities are normalized by their maximal values over the whole domain. The list of them formatted as (time in $10^3$~yr: $p_\text{max}$ in $10^{-11}$~cgs units, $\rho_\text{max}$ in 
  $m_\text{p}$~cm$^{-3}$) is following: 
  (40: 100, 4.3), 
  (45: 74, 4.9),
  (50: 39,  9.8), 
  (55: 31.5, 26),  
  (60: 28,  42), 
  (65: 25,  44),  
  (70: 23, 51),
  (75: 21, 55), 
  (80: 19, 56), 
  (85: 17, 62), 
  (90: 15, 67), 
  (95: 14, 71),
  (100: 15, 75),
  (105: 15, 76), 
  (110: 15, 79),  
  (115: 14.5, 86),  
  (120: 16, 95),  
  (125: 14, 87),  
  (130: 16, 83), 
  (135: 14, 78), 
  (140: 15, 83),  
  (145: 16, 90),  
  (150: 15, 95),  
  (155: 16, 105), 
  (160: 14, 100)}
            
\end{figure*}

We illustrate this with series of density and pressure fields of the run E3T-XY9 in Fig.~\ref{fig:PRhoSeriesFront}.

As one can notice, the shell position matches well the 1D simulations (compare with Fig.~\ref{fig:StructC}), although its compression is 
underestimated due to lower resolution and higher numerical viscosity. While the shell is being formed and accumulates mass, it moves under its inertia. The hot bubble expands into the quickly
developed pressure gap pushing forward a region of high thermal pressure and increased ram pressure of the cooling fluid swept up (matter velocity maxima are situated behind the shell). 

As the hot gas thermal pressure propagates as sound waves, it is subjected to numerical anisotropy of low-mach characteristics inherent to general-purpose Godunov schemes: the flux estimation is based upon 
a superposition of two solutions of the Riemann problem along grid directions, a net numerical error increases as relative weights of those 1D-solutions becomes equal, and therefore effective signal 
propagation speeds differ for characteristics that are close to grid lines and for those considerably inclined.

This effect is stronger for weak discontinuities and sound waves. On Cartesian grids they propagate somewhat faster along the $45^\circ$-inclined direction. 
This explains the fact that the first bends appear around domain 
bisectors on static meshes. Finally, one can notice that the high pressure front, catching up the shell from inside, is `cogged' (see Fig.~\ref{fig:PRhoSeriesFront}a). 

And it is these cogs that ripples the thin shell inner boundary. A high momentum flux density appears focused at the leading
vortices of cogs (because of refraction at the termination shock, or -- more precisely -- a  rear transition region which stands on its place in low-resolution simulations). 
This causes the fluid elements behind them to gain more mass and velocity, and then to penetrate the dense sheet producing the leading arcs. 
Similarly, the fluid before the lagging vortices gains less and consequently sinks in the bubble producing the spikes, suffering greater deceleration and enhancing the termination shock. 
This process also quickly smears the density fluctuations that has been accumulated within the shell before the gap closure. 

After the pressure gap fills up and the hot gas pressure profiles flatten (Fig.~\ref{fig:StructC}b,\ref{fig:PRhoSeriesFront}a), the acceleration saturates and the generated structure continues to expand 
ballistically until the termination shock, reflected from the centre, returns to the shell and gives an additional kick to it, enhancing the bending amplitudes impulsively 
(it is seen in videos available online). 

As the instability is triggered by a transient acceleration of the dense fluid by the light one, it is relative to the Rayleigh--Taylor instability, and to denote it we introduce a special acronym
ARTSI -- an Accelerated Radiative Thin Shell Instability (one can also treat the letters RT there as alluding to the Rayleigh--Taylor mechanism).

\begin{figure*}
   \includegraphics[width=0.95\linewidth]{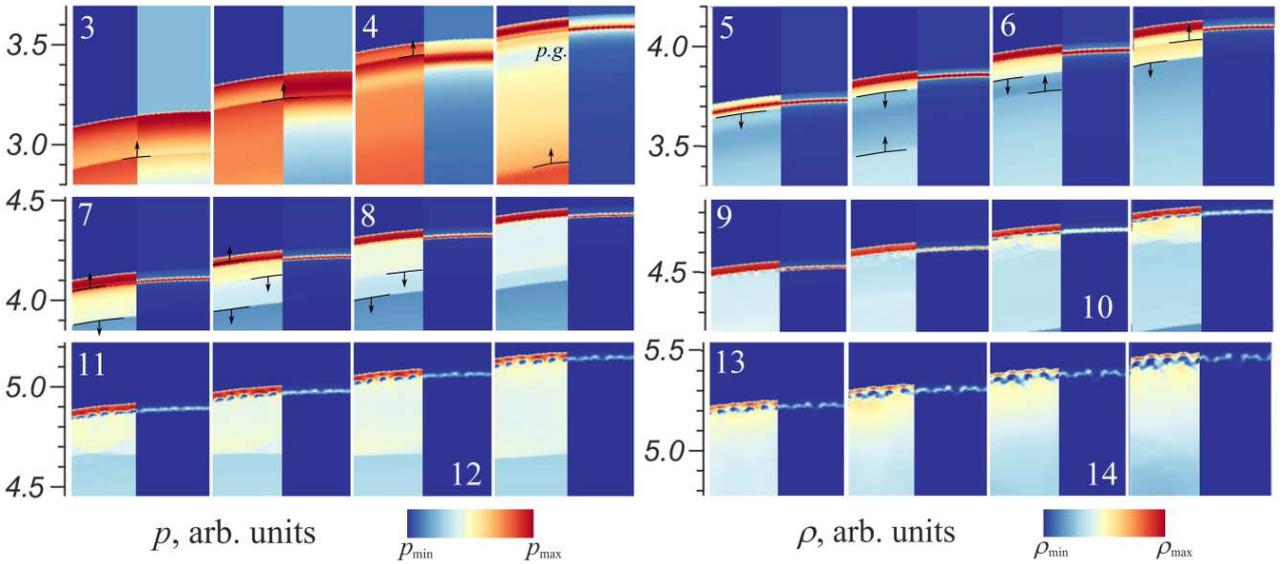}
  \caption{\label{fig:PRhoSeriesFlash} Series of cuts of pressure (left halves of cuts) and density (right halves) fields of the E4T-XY12A run illustrating the shell formation and the  
  instability development. Right boundaries of each cut are aligned to the domain bisector (i.e. $45^\circ$ inclined to both axes), scales to the left are radii from the centre in $10^{18}$~cm, numbers 
  within the cuts indicate time in $10^3$~yr. Black lines and arrows indicate fronts of inner shocks and directions of their propagation. The letters $p.g.$ label the pressure gap. 
  Pressures and densities are normalized by their maximal values over the whole domain. The list of them formatted as (time in $10^3$~yr: $p_\text{max}$ 
  in $10^{-11}$~cgs units, $\rho_\text{max}$ in $m_\text{p}$~cm$^{-3}$) is following: 
  (3.0: 24.3, 4.03),
  (3.5: 16.0, 4.42),
  (4.0: 10.0, 8.05),
(4.5: 9.19, 19.3),
(5.0: 12.4, 46.3),
(5.5: 10.1, 52.4),
(6.0: 8.41, 56.2),
(6.5: 7.32, 71.2),
(7.0: 6.54, 92.5),
(7.5: 6.75, 103),
(8.0: 6.14, 116),
(8.5: 5.58, 137),
(9.0: 5.21, 155),
(9.5: 5.17, 194),
(10.0: 4.94, 231),
(10.5: 4.75, 333),
(11.0: 4.51, 444),
(11.5: 4.24, 446),
(12.0: 4.11, 587),
(12.5: 4.04, 601),
(13.0: 4.21, 646),
(13.5: 4.13, 533),
(14.0: 4.28, 632),
(14.5: 3.95, 705). }
\end{figure*}

Somewhat different picture will show up if one increases spatial resolution. In our experiments it is realized E4-E5 models with similar (or even higher) numbers of cells, but with more than 10 times shorter 
overall span. Corresponding profiles are presented in Fig.~\ref{fig:PRhoSeriesFlash}. For better understanding of evolutions of the flow some internal wave-frons are denoted with thin black lines and arrows indicating
direction of their propagation.

At 3000~yr the cooling is already  setting on, and also one of initial explosion reverberation waves is seen behind the forward shock. By 4 thousand years it merges with the post-shock fluid into a single 
dense shell. Then is partially reflected back, partially passes through the shell and proceeds forward to coalesce with the forward shock. In picture corresponding to 4.5 thousand years the pressure gap is
denoted as `p.g.' and another reverberation wave appears near the lower border. The latter is analogous to the DS3-shell discussed in subsection~\ref{sec:collapse}, or, in other words, it represents matter 
raked by the strong pressure gradient. 

Between 6 and 6.5 thousand years this wave crosses the reverse shock, at 7 strikes the dense shell which by 7.5 results in one more reverse shock reflected into the rear cooling region and another internal shock 
transmitted into the forward one. This strike provides conditions for both Richtmyer--Meshkov instability and the NTSI. However, development of the latter is hindered by quick growth of space between the transmitted 
and reflected shocks which makes the pixel-scale cog-like surface perturbations small in relative amplitude. Another reason is that those perturbations are not in resonance 
(i.e. concavities of one surface do not coincide with convexities of its opposite counterpart). 

Nevertheless, whatever the instability is more efficient it does not affect the dense shell significantly at that time. Visible changes begin when the shell core cools enough so a pressure minimum begins to develop 
within it, similar to Fig.\ref{fig:Fine}a. As the shell is not strictly homogeneous (due to numerical noise which might be enhanced by instabilities taken place earlier\footnote{To test the significance of 
reverberations we have also performed a run of the E4T-XY12A model but with a 10 times reduced $R_0$: there are no echo-waves striking the shell, however the clumpization occurs at the same time in the same way}) 
the thermal instability leads to its clumpization:
in the pressure maps it is seen as separation of the shell into bluish cold clouds. Between these clumps the hot gas from the forward cooling region breaks through into the rear one (additional diverging internal shocks
emerge at spots of its penetration). This is exactly what should be expected from contradirectionality of strong gradients of pressure and density at the forward face of the shell which is therefore Rayleigh--Taylor 
unstable. When the gradients grow high enough and supporting force drops, the rearrangement of the cold-dense and hot-rarefied fluids begins. 
Further we will denote this as FFRT -- the forward face Rayleigh--Taylor instability.

When it finally saturates, the flow somewhat resembles Fig.~\ref{fig:LateStraka}a (separate dense cores in front and low-density `curling smoke' behind them). 
However as in E4T-XY12A typical scales of the flow and numerical noise are smaller than in E3T-XY10, a larger number of fragments is produced, hence they are lighter and decelerate quicker so that
their final dispersion is narrower than in Fig.~\ref{fig:LateStraka}a.

\subsubsection{Physical perturbations on the polar mesh: interplay of Vishniac and Rayleigh--Taylor instabilities}

\begin{figure}
   \includegraphics[width=0.95\linewidth]{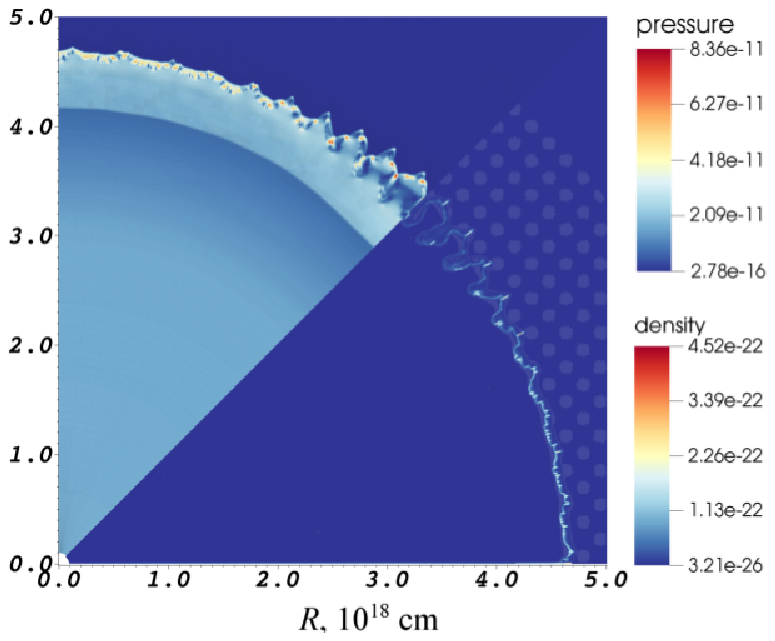}
  \caption{\label{fig:RphPert21} Global picture of the well developed bending and fragmentation due to the RCS passing through density fluctuations when simulated in the E4T-Rph12AD run on the polar mesh. 
                                 In the upper-left half is the pressure field (in cgs units), in the lower-down one -- the density field (in g~cm$^{-3}$). The picture corresponds to 9.5~thousand years after the explosion.}
\end{figure}

Although the RCS flow has been proven unstable in simulations even due to numerical noise only, it is even more important to investigate its response to physical 
perturbations. In the real ISM there is a large variety of density, temperature, velocity and magnetic field fluctuations of different magnitudes and spatial scales, so approaches for proper modelling of interaction of 
SNR-shocks with them constitute a separate kind of art in computational astrophysics \citep[see e.g. recent papers of][and references therein]{KimOstriker14, Korolev15, Martizzi15}.   

In this study we have restricted ourselves with 2D-periodic density fluctuations of wavelength of order 0.1~pc. As we need to resolve the forward cooling region, the short-span
explosion model E4T has been chosen, and the adaptive polar mesh -- to get rid of the cog-like numerical noise discussed above (see E4T-Rph12AD in Table~\ref{tab:models} for details).

\begin{figure*}
   \includegraphics[width=0.95\linewidth]{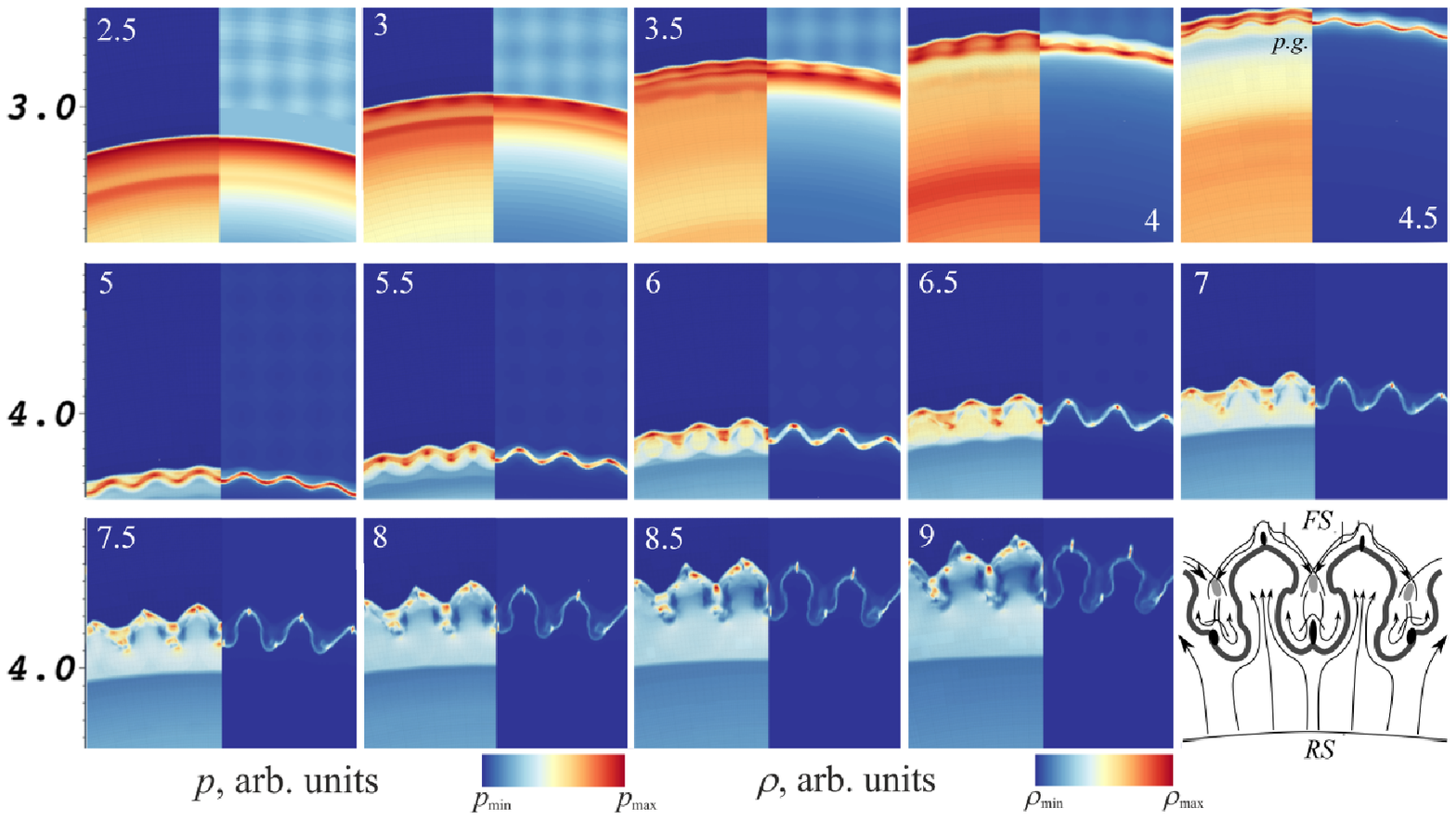}
  \caption{\label{fig:RphPert} Series of cuts of pressure (left halves of cuts) and density (right halves) fields of the E4T-Rph12AD run illustrating the shell formation and development of instabilities near the computational domain bisector. Mid-lines 
  of each cut are aligned to the domain bisector (i.e. $45^\circ$ inclined to both axes), scales to the left are radii from the centre in $10^{18}$~cm, numbers within the cuts indicate time in $10^3$~yr. A cartoon in the bottom left corner
  illustrates positions of the forward (FS) and reverse shocks (RS), corrugated dense shell and clumps as well as flow-lines of matter.
  Pressures and densities are normalized by their maximal values over the whole domain. The list of them formatted as (time in $10^3$~yr: $p_\text{max}$ in $10^{-11}$~cgs units, $\rho_\text{max}$ in 
  $m_\text{p}$~cm$^{-3}$) is following:
  (2.5: 33.3, 3.9),
  (3.0: 24.6, 4.5),
  (3.5: 16.4, 5.8),
  (4.0: 10.9, 11),
(4.5: 10.0, 26.9),
(5.0: 11.8, 40.2),
(5.5: 11.1, 55.4),
(6.0: 10.0, 64.8),
(6.5: 8.5, 105),
(7.0: 8.4, 151),
(7.5: 8.1, 165),
(8.0: 8.2, 182),
(8.5: 7.8, 169),
(9.0: 7.9, 180). }
\end{figure*}

The fluctuations were placed just beyond the radius of the cooling onset and described with following distribution:
\begin{equation}
  \frac{\delta\rho}{\rho_0}=\left\{ \begin{array}{ll}
                                    0.1\sin(kx)\sin(ky), & R\in[3,6]\times10^{18}\text{ cm}, \\
                                    0,  & \text{otherwise}
                            \end{array}  
             \right.
\end{equation}
with $k=2\times10^{-17}$~cm$^{-1}$ and temperature was adjusted to maintain uniform pressure. 

They correspond to a sum of two equal mutually orthogonal plane waves with wavelength $\approx0.07$~pc inclined by $45^\circ$ to the domain boundaries. 
This differs our study from that of \citet{Blondin98}, where density varied only with polar angle and not with radius, and the perturbations comprised in fact a fan of radial rays (or conic surfaces in the 3D-space) 
rather than the alternating clumpy structure. 
In our setup, the shock encountered both fan-like ordered perturbations (along the bisecting line) and those in a staggered order (along borders of the quadrant, see below). This results in a picture
of bending and clumpization shown in Fig.~\ref{fig:RphPert21}.

The difference between the fan-like and staggered perturbation patterns is obvious. Along the bisecting line some parts of the shell for a long time encounter predominantly a density excess  
while their neighbours -- its deficit, and this strongly enhances the bending. The clumps, into which the shell subsequently breaks up, are produced in vertices of the bends, nearly two clumps per one ray of increased density.
A quantity of the clumps is not large, which means that they gain more mass and velocity.

Contrary, near the borders the column density of matter to be swept up is distributed rather uniform with azimuth, which suppresses corrugations (wavelengths are longer amplitudes are smaller) and produces more clumps
which are less massive and have smaller velocity dispersion. 

Fig.~\ref{fig:RphPert} illustrates the instability development in the middle of the quadrant. First, it is seen, that perturbations captured at the earliest stages of the shell formation determine
where leading and lagging arcs will rise and then expand nearly radially. The fluctuations swept up later modify this pattern but not radically (they can not e.g. turn maxima 
into minima). 

Secondly, when the bends of the shock become visible, Vishniac instabilities (caused by refraction of flow lines at oblique shock and their `focusing' behind concavities) start to operate. 
The first spots of high density in the shell appear in the trailing arcs (where the shock encounters increased density), 
however as there is the rippled reverse shock following the shell closely, the leading arcs as well begin to fill-up with matter (and gain momentum). This is direct evidence for the NTSI.

But this `success' of the NTSI is in fact limited. First, the reverse shock departs from the dense shell thus the bends become smaller with respect to the two-shock bounded layer. 
Secondly, they are quickly smoothed out which weakens their positive feedback to the ripples of the forward shock and the dense shell. 

Third (and perhaps the main), such vigorous NTSI manifestation seems to be just a consequence
of the fan-like fluctuation pattern \citep[like in][or in our simulations along the bisecting line]{Blondin98}. In contrast, as it is seen from Fig.~\ref{fig:RphPert21}, when the flow encounters 
perturbations in the staggered order (near the borders) their impacts of alternating sign mutually damp each other considerably. This suppresses development of bends with wavelengths near the pre-shock fluctuations
spatial period. Longer bending waves and a short-scale `jitter' still can be excited, but yet not very efficiently.

One way or another, by 7 thousand years in E4T-Rph12AD there is no large room left for the NTSI to affect the dense shell (especially near the borders), but the latter continues to evolve. 
In the related model E4T-XY12A, as it is mentioned in previous subsection, at this time the shell suffers a certain kick by expanding hot interiors from below and a bit later quickly runs out of pressure 
(due to the losses grow quickly with density), breaks up in a number of drops which then sink in the forward cooling region. Something similar happens to the E4T-Rph12AD flow as well. 

Near the borders all is almost the same: certain growth of pressure in the rear cooling region due to the pressure gap closing, fragmentation into small dense cores and rearrangement of cold-dense 
and hot-rarefied fractions according to the Rayleigh--Taylor instability. 

Along the bisector the situation is more complicated. The forward cooling region is distorted and non-uniform. The advancing arcs 
(and their incipient dense cores) drive bow shocks along which matter flows to the trailing arcs (its flow is terminated there and hot-spots of increased pressure and density are formed). 
Each bow shock is followed by a rarefaction wave lacking both density and pressure (one can see this, e.g. in the $6.5\times10^3$~yr cut as blue low-pressure triangles behind the dark-red bow shocks). 

The pressure at the dense shell rear face is rather uniform, but at the forward face those it alternates considerably, so while some parts of the shell fall in the low-pressure rarefaction waves others are pushed
deeper into the rear cooling region by hot dense spots created by interference of the bow shocks. The dense shell slopes obtain an s-looking shape. 
One can also treat this as Kelvin--Helmholtz instability of an interface between two oppositely directed streams (one above the shell and another -- below it) triggered by inhomogeneities of pressure. 

At the same time, in the vertices of the bends (both leading and trailing) the hot light post-shock fluid tends to flow around the clumps of the cold dense one 
(due to the clumpization and mass concentration by NTSI, its density is even higher than in the case of E4T-XY12A). This instability of the Rayleigh--Taylor kind seems common for all discussed RCSs with resolved 
forward cooling region.

Finally, after crossing the domain of fluctuations, the flow resembles that of Fig.~\ref{fig:LateStraka}a, but with already mentioned distinctions between the 
mid-sector and borders (smaller number of denser cores protruding 2-3 times farther with respect to the shell average).

\section{Discussion}\label{sec:disc}

In this section we are going to describe and discuss briefly key astrophysical and computational consequences of the described effects.

\subsection{Physical vs. numerical effects.}

Let us summarize and classify the instabilities found in this study. 

In 1D-simulations the thermal instability is the only of physical origin. It also gives rise to the `induced fragmentation' which is numerical and is triggered (in our simulations) either by rudimentary 
internal waves caused by too large energy release domains (which can be easily remedied by setting a more proper initial condition) or by the mesh-refinement noise (inherent to computational method, therefore unavoidable
and requiring special caution during interpreting of results).

In multidimensional simulations besides the catastrophic cooling there are different instabilities appear, depending on mesh type and resolution. First of all, there is the `carbuncle' instability which is numerical.
Several physical instabilities can be excited as well due to distortion of characteristic surfaces. 

If the forward cooling region is resolved, than the shell forward face is always Rayleigh--Taylor unstable and its matter tends to switch position with the post-shock fluid in front of it with lower density but higher pressure.
This instability seems rather general and arises independently on whether perturbations are numerical or physical. 

Oppositely, the NTSI which is generally expected in the thin shells, depends on this strongly. 
It can not be excited by numerical noise as the latter acts on short length-scales where the scheme viscosity is high and pixel-sized perturbations are rather small compared to distance between forward and reverse shocks. 
Thus, it is always physical, but in order to be excited efficiently it requires a specially tuned perturbation arrangement, namely a fan-like pattern (transverse phases of fluctuations should not vary 
significantly along paths travelled by shock elements), otherwise it may be damped as well. Finally, the reverse shock departs from the shell and becomes smoother which hampers further development the NTSI and the latter 
gives way to other instabilities.

If the forward cooling region is not resolved (like e.g. in E1T-RZ9 or E3T-XY9), then instabilities depending on the two-shock bounded layer structure do not take place. However, there still appears the ARTSI, 
a Rayleigh--Taylor-like instability of the dense shell suffering an impact from its inner side during the re-establishing of pressure balance between the hot adiabatic inner plasma and the cold dense post-shock fluid 
(violated previously by the sudden cooling increase). In structure-resolved simulations the ARSTI is also possible but usually not detectable as other competing instabilities are triggered earlier. In pure hydrodynamical
modelling it can be observed only due to numerical effects like low resolution or high scheme viscosity preventing the fine structure reproduction, however it also could be physical if there were sources of
high enough non-thermal pressure (e.g. magnetic one) in a certain real shell able to keep its width not very thin and compression -- not very high (e.g. near several tens). 

\subsubsection{Excitation by realistic perturbations}

On Cartesian grids, whatever the resolution, the circular-shaped shell inevitably will be broken up into a number of several-pixel wide dense clumps by the noise-triggered ARTSI or the FFRT. 
This is especially pronounced along directions inclined by $45^\circ$ to the grid axes. This seems artificial unless realistic perturbations fortunately coincide in scales and magnitudes with the numerical noise.

It is unlikely that there can exist short-scale pressure perturbations inside the hot bubble or the rear cooling region: if generated they should quickly decay because of high speed of sound, 
unless they themselves are trans- or supersonic ones (e.g. strong initial explosion velocity fluctuations). 
Indeed, as the hot gas expands into the pressure gap and it is accelerated up to supersonic velocities, hence some pressure or density perturbations may survive and freeze into it. 

Additionally, as there are sub-shells colliding (at least the two spontaneously produced, see subsection~\ref{sec:collapse}),
this also may produce density fluctuations in the shell core. Some perturbations might be introduced by MHD plasma instabilities \citep[like those mentioned in][]{Zirakashvili2008,Slane2015} 
both at forward and reverse shocks.  

Also, there always may be `outer' perturbations like ambient matter inhomogeneities. 

\subsubsection{Catastrophic cooling stabilization}

The second important question regards physical constraints on the shell peak compression, and therefore its cooling time, density and pressure gradients at its forward face as well as accelerative ability 
of the pressure gap. 
First, one should be careful with low-temperature cooling treatment. It has been already shown in subsection~\ref{sec:kinematics}, that a cooling cut-off can modify the shock dynamics and the shell structure. 

Additionally, the catastrophic cooling might be slowed down by magnetic field pressure which should prevent contraction. In recent paper of 
\citet{PetrukKuzyoBeshley} this question has been addressed, and it has been shown, by means of high-resolution spherically symmetric ideal MHD-simulations, that presence of 10~$\mu$G tangential  
magnetic field before the shock can saturate the shell compression at a level as low as 10. However, like in \citet{Blondin98}, their cooling function is likely cut off at $T=10^4$~K which means that the 
shell regains `elasticity' and begins to spread much quicker (than in our simulations) even without magnetic fields. 

It is required to clear, what might happen to the shell if it for some reason were overcooled well below the hydrogen recombination temperature. 
Obviously, neutrals are insensitive either to Coulomb collisions or  to collective interaction or magnetic fields, so they can drift through the hot ionized 
fluid (due to velocity difference between the cooling regions and the dense core, see Fig.~\ref{fig:Fine}) until become ionized collisionally. This path should be 
rather long because collisional ionization rates (in both cooling regions) are several (8-10) orders of magnitude lower than those of Coulomb collisions or plasma frequencies (which are responsible for hydrodynamics of the 
ionized fraction). 

As the ionization is realized via direct collisions, its rate should be comparable to the cooling rate (the latter is caused mainly by collisional excitation of lines), in particular it is interesting to
estimate which fraction of neutrals could overtake the shocks \citep[see e.g. recent studies of effects of neutrals and partial ionization on the forward shock structure in][]{MorlinoBlasi, Ohira}. 

One can note that direct electron-neutral collisions could provide either an extra cooling (when velocities of the neutrals do not 
exceed bulk velocity of plasma considerably, they just extract energy from pool of electrons), or additional momentum deposition (if the neutrals penetrate into a much slower plasma).
The latter mechanism, obviously, could sustain pressure balance within the two-shock bounded shell even if its core becomes mainly neutral. 
It is also interesting to check, whether this mechanism (and corresponding ionization fronts) is unstable or overstable. 

\subsection{Recommendations for modelling of the RCS thin shells.}

First, it is important to determine which mesh spacing is required to reproduce correctly the shell and the cooling regions with a given cooling function and magnetic fields, avoiding considerable scheme viscosity 
(i.e. the shell being only few pixels wide, which also means relatively strong numerical noise). From our estimations (see e.g. equation~\ref{eq:limdr}), for example 512-1024 Cartesian cells per dimension seems rather
poor choice for real SNRs. At least two more orders of magnitude are required\footnote{Although some authors \citep[like e.g.][]{KimOstriker14} who obtained the ARTSI in their coarse-grid multiD Cartesian simulations conclude that
it is unimportant for the physics of interest and one in general can still rely upon low-resolution models.}.

The bending and fragmentation of the RCS in unperturbed media observed on Cartesian grids result from a strong amplification of numerical noise by the physical instabilities. 
The FFRT seems to appear in almost all Cartesian simulations with resolved forward cooling region at any stage after the dense shell formation, the ARTSI is transient.
Thus one should also check whether the computation domain covers regions of transient acceleration or pulsations. 

These estimations can be made through a detailed analysis of results of simple preliminary 
1D-calculations. The temperature and density profiles may be indicative of the thermal instability, while pressure and acceleration curves -- of the ARTSI. 

If it is known definitely that there should be no physical conditions for ARTSI development in a certain real RCS, then to avoid its artificial numerical excitation and not to refuse using simple Cartesian grids,
one can perform 1D-calculations somewhat beyond the instability zone (this would allow a better spatial resolution as well) and then remap them on to the multidimensional domain.

In other cases (when regions of acceleration and/or pulsations do lie within the domain of interest), as the numerical noise has no astrophysical meaning and also depends on mesh type, resolution and use of refinement,
then its growth is undesirable and, therefore it is strongly recommended to choose as `noise-free' mesh as possible, i.e. the most completely aligned to the flow lines. 
On Cartesian grids the instabilities will always appear unless the shock is strictly plane-parallel 
(which, nevertheless, also may be the case when one studies a fine structure of a small localised part of the RCS shell with high resolution).

In our case it was sufficient to use polar grids. However, nowadays, of the highest interest are complex multidimensional flows, generally 
inconsistent with either spherical or planar symmetry, and probably -- with exception of some specific cases -- it is unlikely to choose a proper Eulerian grid for the most of them. 
Actually, even a study of the thin shell instabilities with realistic perturbations (which are neither quadratic nor of a polar mesh cell shape) is one of such cases: 
it will always suffer from the numerical noise to a certain degree.

Even the AMR can not be a universal remedy, since its algorithms are usually triggered by hydrodynamical variables nonuniformities only after they have already grown high enough, 
and therefore can not prevent initial rise of numerical perturbations. 
However, as the instabilities develop near the shocks, where the resolution is kept high, then it seems possible (at least in expectations) by using a `brute force' of the highest refinement level available to 
clear the numerical noise off the scales of interest, i.e. the shell itself and wavelengths of physical perturbations.  

On the other hand, if one simulates a physically one-dimensional flow on a multiD aligned mesh (e.g. a spherical shock on a polar grid), 
the numerical `carbuncle' and `odd-even' instabilities may come out (as they really do in our simulations near the axes) and make the situation even worse than the ARTSI and FFRT, 
so one may need to experiment with different Riemann-solvers as well.   

\subsubsection{Problems for future studies}

Further studies of the RCS instabilities should overcome the limitations of this work.

Inclusion of correct low-temperature physics of heating-cooling and ionization state (as well as their effects on to medium resistivity) should be one of primary 
objectives. Taking into account the time-dependent of ionization (at least of hydrogen) and cooling like in \citet{PlutoRadCool} is also desirable, especially in high resolution simulations.
Correct account of variable electron fraction should help to avoid the low-temperature cooling constraining by a voluntary prescription.  
It is also required to include magnetic fields as they limit the shell compression.

Additionally as there are several different characteristic time- and length-scales, it would be better in general to perform modelling in 3D and as close to real flows of interest as possible to obtain realistic instability patterns, 
triggered by the most realistic physical perturbations of velocity, pressure, magnetic field or ambient density, which are likely to affect spectra of the shell fragments.

\section{Conclusions}\label{sec:conc}

In this paper, by means of numerical experiments the reliability of which is ensured by consistence of results of simulations with three independent well-developed codes, 
we show that during the transition from the adiabatic regime to the radiative cooling, the multidimensional shock flow is distorted and fragmented not only and not so much due to the Non-linear Thin Shell Instability 
of Vishniac, but due to the Rayleigh--Taylor-kind instabilities connected either to the thin dense shell transient acceleration during the dense shell collapse and subsequent pressure equilibration with hot interiors
or to its potentially unstable inner structure.
We have designated them as the Accelerated Radiative Thin Shell Instability (ARTSI) and the Forward-Face Rayleigh--Taylor (FFRT) instability, respectively. 

It is found that both are very susceptible to the numerical noise superimposed by Cartesian grids on to the curvilinear surfaces of shells and shocks. The NTSI in contrast appears insensitive not only to the numerical noise 
(on the pixel-scales it is suppressed by the scheme viscosity, misadjustment of the noise ripples of the shock surfaces and by growth of distance between them), but also to rather strong physical perturbations 
(10 per cent of the ambient density) if they are not arranged close to a certain fan-looking pattern. Even under the most favourable conditions, the NTSI growth can be quickly terminated because of the reverse shock 
departure and smoothening. Nevertheless, the NTSI-produced clumps serve as strong non-linear perturbations for further development of thermal and Rayleigh--Taylor instabilities. 

Under certain conditions, this may lead to fragmentation of the RCS dense shell at the earliest stages of its existence.

Another important part of our work is the high-resolution one-dimensional study the dense shell formation, dynamics and structure. It is found that all of them considerably depend on the low-temperature
cooling treatment. The structure and sub-structure are also sensitive to the grid resolution and a scheme viscosity contribution to the pressure balance. Some possible artefacts of numerical methods 
and/or statement of initial conditions which may give rise to additional dense shells are also discussed.

These observations somewhat improve our understanding of evolution of the non-stationary radiative shocks (e.g. in SNRs) and also set new requirements on further perspective studies of astrophysical flows of such kind. 

\section*{Acknowledgements}

We thank Prof. Sam~A.~E.~G.~Falle, Prof.~A.~Mignone, P.~Creasey, K.~Kifonidis and K.~V.~Krasnobaev for helpful discussions and references. 

This work is supported
by Russian Science Foundation grant 14-12-00203. 
SIG thanks the `Dynasty' foundation for support of development of the \small{FRONT3D} code.
SIB also thanks SNF (SCOPES) grant IZ73Z0-152485 for support of work on SNR cooling. 
The software used in this work was in part developed by the DOE NNSA-ASC OASCR Flash Center
at the University of Chicago.

\bibliography{RShDB4L}
\bibliographystyle{mnras}

\appendix

\begin{center}
  \bfseries {\Large APPENDICES}
\end{center}

\section{Exact implicit cooling procedure}\label{app:CoolTown}

Here we briefly describe the exact cooling procedure used in the \small{FRONT3D} code for temperature calculations,
 based upon the algorithm proposed in~\citet{Townsend_ApJS_2009}. It improves precision of cooling
calculations and removes possible ambiguities of solution. 

The energy equation (\ref{eq:rho_e_cool}) is subdivided into an advection sub-step (with its right hand side neglected) and a cooling sub-step, treating only radiative cooling of stationary matter 
($\rho$, $n_{\text{e}}$, $n_{\text{H}}$ are considered constant) with equation~(\ref{eq:rho_e_cool}) reduced to:
\begin{equation}\label{eq:T_cool}
  \rho\partial_te=\frac{n_{\text{H}}k_{\text{B}}}{(\gamma-1)\mu}\partial_t T=-\Lambda(T)n_{\text{e}}n_{\text{H}}.
\end{equation}
The formal solution for the time-step from $t$ to $t+\Delta t$ reads:
\begin{equation}
  \int\limits_{T(t)}^{T(t+\Delta t)}\frac{\text{d}T'}{\Lambda(T')}=-\frac{(\gamma-1)\mu n_{\text{e}}}{k_{\text{B}}}\Delta t.
  \label{eq:T_cool_sol}
\end{equation}
Following \citet{Townsend_ApJS_2009} we define
\begin{equation}
  Y(T)\equiv \frac{\Lambda(T_{\text{ref}})}{T_{\text{ref}}}
  \int\limits_{T}^{T_{\text{ref}}}\frac{\text{d}T'}{\Lambda(T')},
  \label{eq:Y_def}
\end{equation}
where $T_{\text{ref}}$ is some reference temperature, and then rewrite the solution of Eq.~(\ref{eq:T_cool_sol}) in terms of $Y(T)$:
\begin{equation}
  T(t+\Delta t)=Y^{-1}\left[Y(T(t))+\frac{\Lambda(T_{\text{ref}})}{T_{\text{ref}}}\frac{(\gamma-1)\mu n_{\text{e}}}{k_{\text{B}}}\Delta t\right].
  \label{eq:townsend_cooling}
\end{equation}

The functions $Y(T)$ and $Y^{-1}(Y)$ are pre-calculated and tabulated over the whole temperature range of $\Lambda(T)$ (coefficients of interpolation and extrapolation are also provided).
As $Y(T)$ is strictly monotonous, then $Y^{-1}(Y)$ should give a unique solution for Eq.~(\ref{eq:townsend_cooling}) for any arbitrary $\Delta t$, and this is the main advantage of the method 
before simple inversion of the non-monotonous $\Lambda(T)$ by Newton--Raphson solvers.

\section{The instability evolution in different codes}\label{app:InstEvol}

Here we present series of density profile pictures obtained by means of various codes to illustrate the thin shell formation and development (or decay) of the ARTSI in different physical 
and computational conditions. They are also available online and through our group's web site \url{http://dau.itep.ru/sn/radshocksnr/} as well as other high-resolution pictures and videos.

\subsection{The ARTSI on fixed regular Cartesian grids}

\begin{figure*}
   \includegraphics[width=0.95\linewidth]{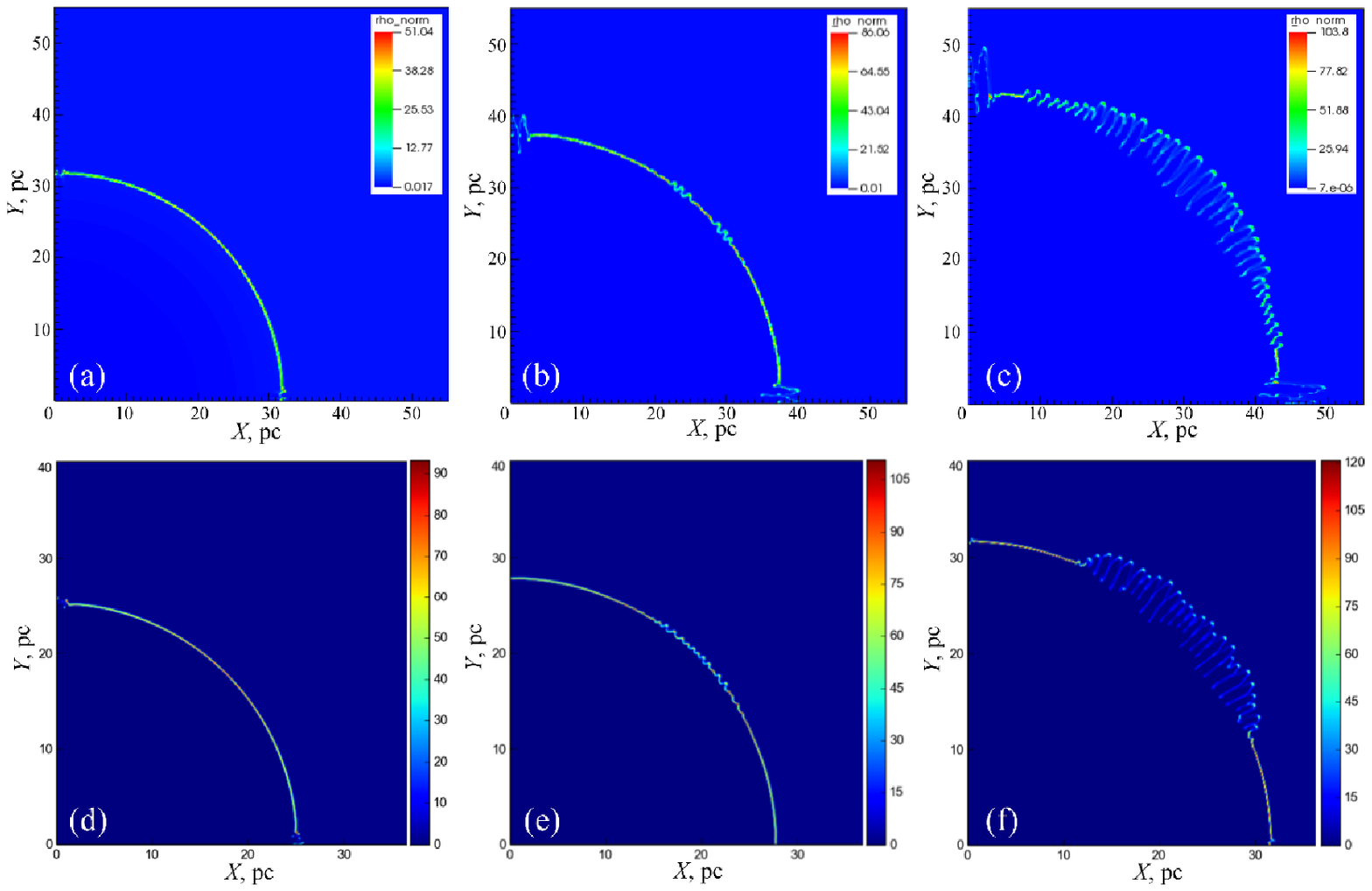}
  \caption{\label{fig:StatMesh} Bending dynamics on fixed meshes. E3T-XY9 simulation density profiles (in $m_\text{p}$~cm$^{-3}$) output by {\small FRONT3D}: 
  (a) -- the shell formation and `carbuncle', (b) -- the first ARTSI bends, (c) -- late well developed ARTSI bends. 
  (d),(e),(f)  -- the same but for E6F-XY9 and {\small PLUTO4}, respectively.}
\end{figure*}

Fig.~\ref{fig:StatMesh} displays the ARTSI dynamics in {\small FRONT3D} and {\small PLUTO4} simulations on regular rectangular meshes without AMR, obtained in models E3T-XY9 and E6F-XY9, 
respectively. The three typical stages are present:
\begin{enumerate}
\item the shell is just formed and no perturbations are seen but only a near-axis `carbuncle'; 
\item the first ARTSI ripples arise; 
\item the ARTSI is well developed, especially around the breaks of the discretized shell.
\end{enumerate}

The bending pattern looks rather symmetric with respect to the domain bisector. After some exploration of the E6F-XY9 figures, it can be seen that the bends resemble stretched Rayleigh--Taylor mushrooms. 

\subsection{The ARTSI on Cartesian grids with AMR}

Here we show how the AMR can change the picture of the shell rippling. All figures are taken with the {\small FLASH4} code and the Table cooling function. It is seen, that `mushrooms' appear
in the shocked fluid, collapse into dense droplets due to the thermal instability and somewhat later pierce and ripple the forward shock and the shell. 

The bending pattern seems to have both symmetric and antisymmetric modes.

\begin{figure*}
  \includegraphics[width=0.95\linewidth]{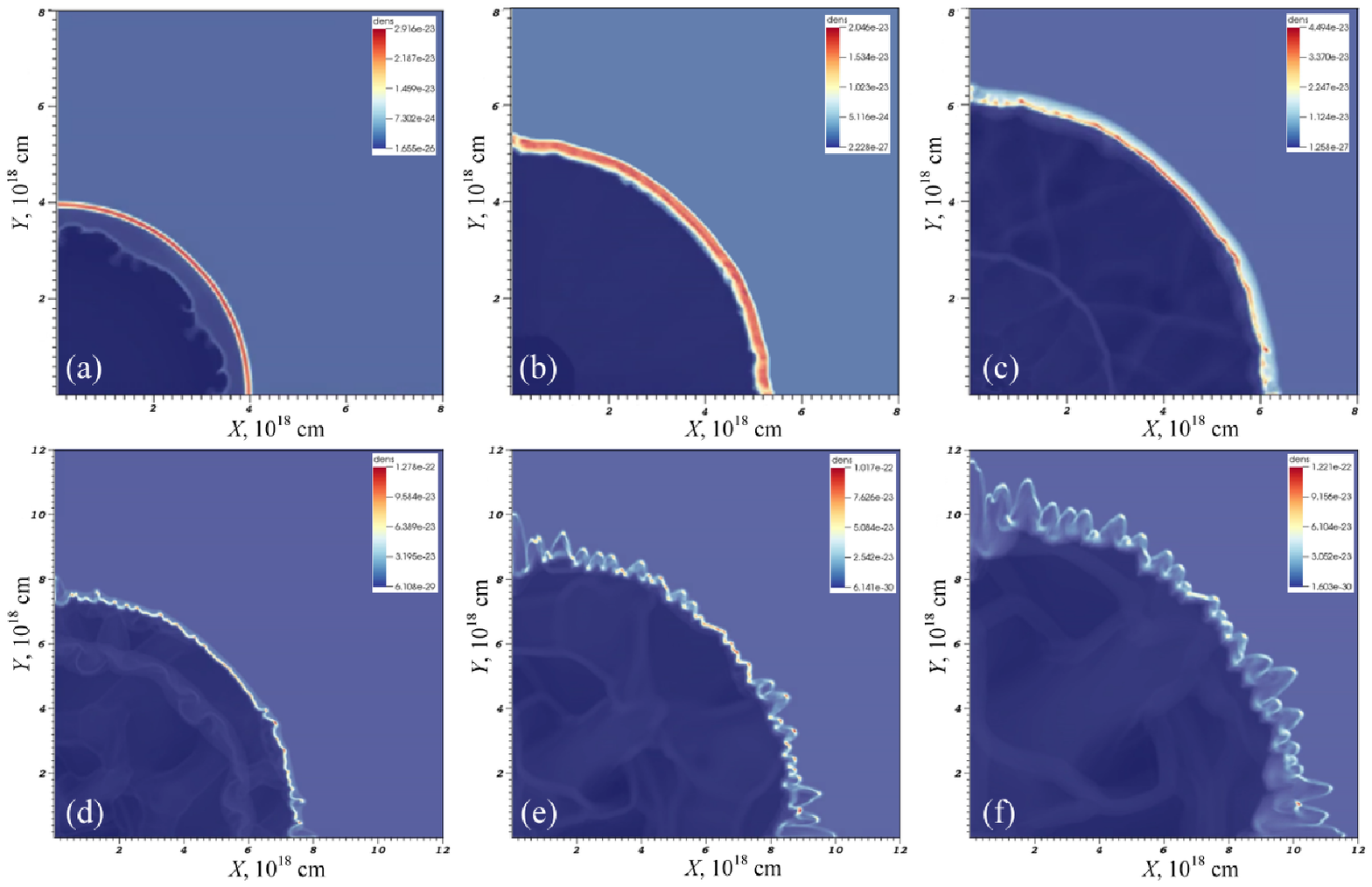}
  \caption{\label{fig:E4TAMR} Bending dynamics for E4T model on AMR mesh by \small{FLASH4.2}. 
  Density profiles (in g~cm$^{-3}$): (a) -- the shell is just formed; (b) -- the first density fluctuations appear within it; (c) -- they `condense' into droplets 
  behind the forward front; (d) -- the droplets overtake the shock and produce the first bends; (e) -- the bends cover the whole shell limb; (f) -- the shell is broken into separate dense cores 
  driving their own bow shocks.}
\end{figure*}

\begin{figure*}
  \includegraphics[width=0.95\linewidth]{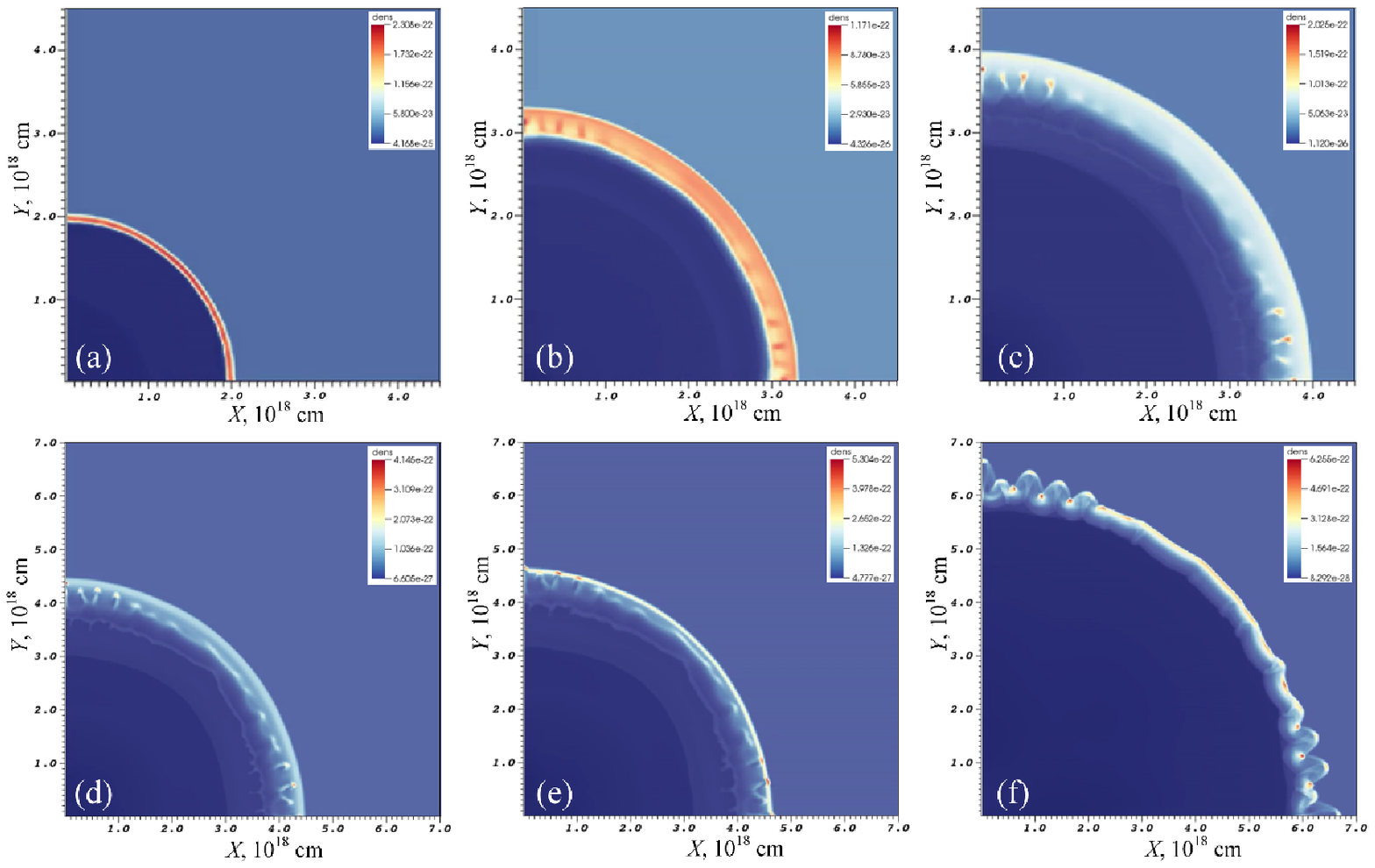}
  \caption{\label{fig:E5TAMR} Bending Dynamics for E5T model (higher ambient density) on AMR mesh by \small{FLASH4.2}. Density profiles (in g~cm$^{-3}$): (a) -- the shell is just formed; (b) -- the first mushroom-shaped density fluctuations appear within it; 
  (c) -- they `condense'  into droplets behind the  forward front; (d) -- the dropls move towards the shock; (e) -- the drops catch up the shock and the shell; (f) -- the dense cores with their bow 
    shocks has gone out of the shell which is still thin, dense and slightly rippled.}
\end{figure*}

\bsp	
\label{lastpage}
\end{document}